\documentclass[%
 reprint,
 amsmath,amssymb,
 aps,pre
]{revtex4-1}
\pdfoutput=1

\usepackage{graphicx}% Include figure files
\usepackage{dcolumn}% Align table columns on decimal point
\usepackage{bm}% bold math
\usepackage[caption=false]{subfig}
\usepackage{color}

\begin{document}

\preprint{APS/123-QED}

\title{Phase behavior of blocky charge lattice polymers:\\Crystals, liquids, sheets, filaments and clusters}

\author{Nicholas A.S. Robichaud}
%  \email{nasr12@mun.ca}

\author{Ivan Saika-Voivod}%
% \email{saika@mun.ca}
\author{Stefan Wallin}
  \email{swallin@mun.ca}
\affiliation{%
 Department of Physics and Physical Oceanography\\
 Memorial University of Newfoundland
}%

\date{\today}% It is always \today, today,
             %  but any date may be explicitly specified

\begin{abstract}
Motivated by the idea that intrinsically disordered proteins (IDPs) condense into liquid-like droplets within cells, we carry out Monte Carlo simulations of a polymer lattice model to study the relationship between charge patterning and phase separation. 
Polymer chains containing neutral, positively charged and negatively charged monomers are placed on a cubic lattice. Only nearest neighbor interactions between charges are considered.
We determine the phase diagram for a systematically varied set of sequences. We observe homogeneous fluids, liquid condensation, cluster phases, filaments, and crystal states. Of the six sequences we study, three form crystals at low temperatures. The other three sequences, which have lower charge densities, instead collapse into gel-like networks or unconnected finite clusters. Longer neutral patches along the sequence sterically limit the size and shape of low-energy structures, which is analogous to the effect of charge or limited valence in attractive colloids. Only one sequence clearly exhibits liquid behavior; this sequence has a reduced tendency to individually fold and crystallize compared to others of similar charge density and draws parallels to real IDP behavior. 

\end{abstract}

%\keywords{Suggested keywords}%Use showkeys class option if keyword
                              %display desired
\maketitle

\section{\label{sec:level1}Introduction}
The self-organization into complex structures at different length scales is one of the hallmarks of biological systems. A well-known example at the molecular level is the folding of proteins into compact  globular three-dimensional structures, driven primarily by hydrogen bonding, hydrophobic and electrostatic interactions~\cite{Dill1990}. Not all proteins fold spontaneously into stable conformations, however. A substantial fraction of the proteins in many genomes are conformationally flexible even under physiological conditions~\cite{Uversky2002, Tompa2005,wright}. Rather than folding into a unique structure, these so-called ``intrinsically disordered proteins'' (IDPs) continuously interconvert between different conformations. IDPs tend to have low-complexity, sometimes repetitive, amino acid sequences~\cite{Jorda2010}. Compared to more ordered proteins, IDPs are depressed in hydrophobic residues and enriched in small, polar and charged amino acid residues, such as glycine, serine, lysine and glutamic acid~\cite{Radivojac2007}. The highly dynamic and flexible nature of IDPs make them ideal for certain biological functions. For example, the ability of IDPs to transiently bind many different ligands and respond quickly to changes in their environment~\cite{Dunker2005}, likely underlie their high prevalence among signalling and cell regulatory proteins~\cite{Zhou2012}. 

This multivalency in IDP interactions also plays a key role in the formation of very large biomolecular assemblies in cells, otherwise known as membrane-less organelles~\cite{fung}. The lack of an enclosing membrane allows these condensates to quickly adapt to changes in the cellular environment and exchange components with their surroundings, making them aptly suited to roles in signalling and regulation~\cite{fung}. Membrane-less organelles can also isolate material from the rest of the fluid cellular environment, allowing the storage or processing of biomolecules like RNA~\cite{brangwynne}. Examples of biomolecular condensates with important cellular functions include nucleoli, the site of ribosomal biogenesis~\cite{ Feric2016}; stress granules, which respond to adverse cellular conditions such as heat stress~\cite{riback2017}; processing bodies (P-bodies), which are important in mRNA processing~\cite{Luo2018}; and germ granules, RNA-storing bodies critical in embryonic development~\cite{brangwynne2009}. All these molecular assemblies are enriched in IDPs~\cite{fung}.

Many membrane-less organelles has been found to form through a liquid-liquid phase separation, i.e., they are condensed liquid-like droplets with higher densities than the surrounding fluid but still retain liquid properties~\cite{shin}. Brangwynne \textit{et al.}~\cite{brangwynne2009} showed that nematode germ granules are roughly spherical in shape, fuse on contact, rapidly dissolve and condense, and exhibit wetting and dripping behavior characteristic of liquids. Nucleoli were shown to be liquid-like droplets which are approximately four orders of magnitude more viscous than water~\cite{Brangwynne2011}. Some other cellular condensates are more gel-like~\cite{shin} and cannot be neatly characterized as either liquid or solid. For all membrane-less organelles, formation is dependent on the conditions of the cell, such as changes in protein or RNA levels, salt or proton concentrations, or temperature~\cite{brangwynne}. For instance,  stress granules formation can be triggered by a change in temperature, thus potentially helping to protect the cell from the debilitating effects of heat shock~\cite{riback2017}. The transition to a liquid-like condensed state can also be a precursor to the formation of pathological, fibrillar structures, such as the case of stress granules in conditions of persistent, severe stress on the cell~\cite{molliex2015}. 

Specific IDPs have been identified as the primary drivers of phase separation for several types of membrane-less organelles, and some IDPs have been shown capable of forming liquid-like droplets on their own~\cite{fung,brangwynne}. Experimental work with IDPs has suggested that the pattern of charged amino acids along the sequence is important for phase separation~\cite{shin,Nott2015,pak2016}. For example, LAF-1, an IDP thought to be important in germ granule formation, contains a region with blocks of oppositely charged aspartic acid and arginine residues. This region with a ``blocky'' charge pattern has been shown to be sufficient to drive the formation of a liquid droplet on its own~\cite{elbaum2015}. Another related germ granule protein, DDX4, also self-assemble and drive liquid droplet formation. A charge scrambled variant of this protein, with the overall amino acid composition maintained, lost its ability to phase separate~\cite{Nott2015}. These findings highlight the importance of both the presence of charged amino acids and their pattern along an IDP sequence in driving phase separation of membrane-less organelles.  Other types of short-range interactions including hydrophobic interactions between aromatic ring structures and dipole-dipole interactions are also thought to play a role~\cite{schmidt2015,brangwynne}. Overall, the physical driving forces of IDP-driven liquid-liquid phase separation need to be elucidated~\cite{shin}.

Attempts have been made at using coarse-grained simulations~\cite{Das2018,Dignon2018, harmon,Nguemaha2018,Xu2019,chan} and field-theoretical approaches~\cite{Lin2016,Danielsen2019} to get insight into the relationship between amino acid sequence and the liquid phase behavior of IDPs. For example, Chan \textit{et al.}~\cite{chan} studied the role of charge patterning in driving phase separation using a one-bead per amino acid lattice model and Monte Carlo simulations. They compared two charge-neutral 50 monomer sequences, one with purely alternating charges, and one with blocks of positive or negative charges arranged asymmetrically throughout the sequence. From simulations at different temperatures and chain concentrations, they observed that the chain with a ``blockier'' charge pattern had a higher tendency to phase separate, supporting the current understanding that blocky charge patterns in IDP sequences promote phase separation.
A strength of Chan~\textit{et~al.}~\cite{chan} is that liquid droplets formed spontaneously in their simulations, which were initialized from random chain configurations. By contrast, in many off-lattice simulations of liquid-liquid phase separation carried out so far, initial configurations are prepared through a (non-equilibrium) procedure in which chains are compressed into a compact, liquid state ~\cite{Dignon2018,Das2018,harmon,Nguemaha2018}.  A limitation of  Chan~\textit{et~al.}~\cite{chan} was that only sequences with exclusively charged residues were considered, whereas  proteins contain a mix of charged and neutral residues. Additionally, the densities of coexisting phases in their simulations could not be well identified, and thus their characterization of the phase behavior was approximate. Their model also highlighted some of the computational challenges with simulating phase separation in large polymer systems. At lower temperatures, their simulations did not reach a stable equilibrium, instead trending towards lower energy states. This suggests that much longer simulations would be required to fully analyze their systems, which might not be computationally feasible. With all this in mind, further work must be done to fully analyze the role of charge patterns along IDP sequences.

In our work, we use a simpler but more general lattice model to study the phase behavior of charged polymers. We use longer Monte Carlo simulations with a broader outlook to examine how charge patterning influences phase separation for six representative sequences. These sequences are 24 monomers in length, contain neutral, positive and negative monomers, and have different charge densities and charge patterns. From our simulations, we look to identify the structural features of the phases that form at different temperatures and densities and construct phase diagrams for these sequences. From this, our work further characterizes the role of sequence-dependent electrostatic interactions of IDPs in the formation of biomolecular condensates in cells.

\section{Methods}

\subsection{Lattice model simulations}
We use lattice model simulations to study the phase behavior of charged, linear polymers, with the aim of modeling the behavior of IDPs. Our model consists of 24-bead chains placed on a cubic lattice, where beads are positively charged, negatively charged, or neutral. Each simulation involves 300 chains. The coordinates of beads are confined to specific lattice sites that are equally spaced throughout the three-dimensional box. The beads along a chain are placed in adjacent lattice sites, such that the bond angles between neighboring beads are either 180$^\circ$ or 90$^\circ$. There is no overlap permitted on a single lattice site. The simulation box is given periodic boundary conditions, allowing us to approximate the behavior of a bulk system, since surfaces are removed.

In our model, only nearest-neighbor interactions are considered  when calculating the system energy $E$. Every site on the cubic lattice has six nearest-neighboring sites. Charges interact in the expected way, with like charges repelling and opposite charges attracting, given an energy of interaction of $+\epsilon$ or $-\epsilon$, respectively. Neighboring beads along a chain are not given an energy of interaction, since this would only add a constant to the energy of the system. Neutral beads and empty sites on the lattice have no interaction energies. Thus, only nearest-neighbor electrostatic interactions between charges that are not adjacent along a chain are included in the model.  If the sequence of a chain with $N$ beads is represented by $s=(s_1,...,s_{N})$, where $s_i=0,+1, \text{or} -1$ is the charge of the $i$th bead along the chain, we can express the total energy of $M$ such chains as
\begin{equation}
    E = \epsilon \sum_{i=1}^{NM}\sum_{j=i+1}^{NM} s_{\alpha_i} s_{\alpha_j} \Delta_{ij} (1- \delta_{\zeta_i,\zeta_j}\delta_{i+1,{j}})\,,
\end{equation}
where the beads in the $M$-chain system have been labelled $1,...,NM$, $\alpha_i = i - N\zeta_i$, $\zeta_i$ is the chain number of bead $i$, $\delta_{ij}$ is the Kronecker delta and $\Delta_{ij}=1$ if beads $i$ and $j$ are neighbors on the lattice, and zero otherwise. The fourth factor of the summand removes contributions from pairs $ij$ adjacent along a chain.

The progression of a simulation can be considered in terms of steps. Every step is an attempt to randomly move a bead or portion of a chain. If the move is determined to be viable, an energy calculation is performed, and the move is accepted or rejected based on the Metropolis Monte Carlo criterion~\cite{chandler}. Our simulations involve over 300 billion of these steps performed consecutively. For the most part, we denote the length of simulations in cycles. A cycle represents $N$ steps, where $N$ is the number of beads in the simulation box (7200 for most of our systems). Measuring simulation length in this way improves the consistency in comparing results from simulations with different numbers of beads. 

\begin{figure}
 	\includegraphics[width=1.0\linewidth]{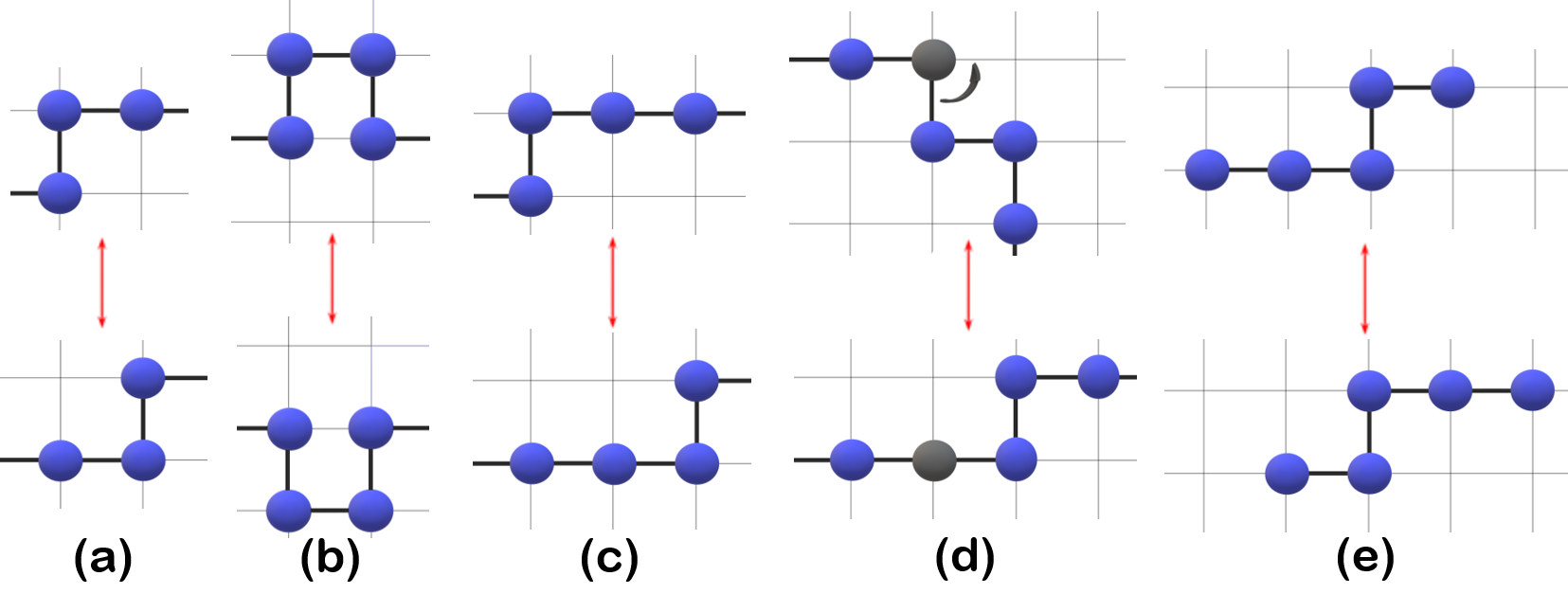}
 	\caption{\label{fig:moves}The types of moves included in the simulations. The one-bead moves are corner flips (a) and rotations of end beads (not shown). Two-bead moves are crankshaft rotations (b) and L-flips (c). Pivots (d) involve rotations or reflections of a portion of a chain around a pivot point, shown in gray. Reptations (e) are sliding movements of an entire chain forward or backward by one bead. The figure shows an example with a five-bead chain; the sequences in our project are 24 beads long.}
\end{figure}

Every step, the simulation randomly picks between four types of moves to conduct: a one-bead move, a two-bead move, a pivot, or a reptation. Once a type of move is picked, a random target bead or chain for that move is chosen, a potential final configuration resulting from that move is generated, an energy calculation is made, and, finally, the move is accepted or rejected. If no viable move is possible with the chosen move type and bead, then that simulation step ends without performing an energy calculation or changing the system configuration. Examples of all possible moves are shown in Fig.~\ref{fig:moves}. There are two types of one-bead and two-bead moves included. One-bead moves include end-bead rotations, where the terminal bead in a chain is rotated to any free lattice site adjacent its neighboring bead in the sequence, and corner flips, which shift an internal bead as shown in Fig.~\ref{fig:moves}a. Two-bead moves include crankshaft rotations and L-flips, which shift the position of two beads when they are found in specific conformations (Fig.~\ref{fig:moves}b and \ref{fig:moves}c). Pivots are performed by first selecting an internal bead along a chain to act as a pivot point. Then, a randomly selected rotation or reflection is attempted. Portions of the chain can be rotated 90$^\circ$, -90$^\circ$, or 180$^\circ$ along the $x$, $y$, or $z$ axis, or reflected along the $x$, $y$, or $z$ axis around the pivot point. Finally, reptations involve a snake-like movement of an entire chain. After a random chain is selected, the simulation attempts to move the first or last bead in the chain to a randomly selected adjacent lattice site. All other beads in the sequence then follow along with this movement, sliding forward or backwards along the initial coordinates of the chain (Fig.~\ref{fig:moves}e). All moves obey detailed balance: they are completely reversible, and the reverse of a move has the same probability of being attempted as the initial move had. Acceptance or rejection of a move is based in part on its associated energy. 

The average acceptance probability for each move depends on the temperature, the system density and the sequence, but in most systems is roughly 40\% for one-bead moves, 20\% for two-bead moves, 30\% for pivots and 10\% for reptations. The moves are not attempted with the same probability. In the final version of the simulation code, 45\% of steps are one-bead move attempts, 25\% are two-bead move attempts, 25\% are pivot attempts and 5\% are reptation attempts. These numbers were adjusted over time to maximize the computational efficiency of the simulation. In doing so, we look for a balance between the number of steps required to reach an equilibrated state and the computational time required for those steps. Larger scale moves like a reptation can reach a stable, low-energy state in a low number of steps, but each step involves costly, slow energy calculations, so swaying the simulation towards other move types increases the overall efficiency.

We investigate the idea of implementing a cluster move into the simulation. This move would translate an entire group of chains that are in contact with each other through the simulation box. Ultimately, we conclude that this type of move is not worth the computational cost. Moving a cluster requires first defining which chains are clustered together, a costly process. The move also creates difficulties with detailed balance, where if the cluster size increases because of the move (if it encounters new chains), the move is not directly reversible; any such moves must be rejected. These problems outweigh any potential benefits of a cluster move. Therefore, in the simulations in this work, a large cluster will only translate through the box through the sum of individual movements of its composing chains. As such, a dense cluster where interior chains have very little conformational freedom will be largely immobile.

Once a viable move is found, the simulation calculates the difference in energy between the initial conformation of the relevant beads and their final conformation. This energy change and the temperature of the simulation are used to determine whether the move will be accepted in accordance with the Metropolis Monte Carlo method~\cite{chandler}.

To ensure that simulations are not biased towards certain conformations, a random system configuration is generated at the beginning of each simulation. This is done by performing one million random moves at infinite temperature, i.e.~where all moves are accepted regardless of energy (though bead overlaps are still forbidden). This creates a random configuration of chains that is used as the initial state for the simulation proper.

\subsection{Observables}
During the simulation, the system energy is calculated periodically and printed to an output file. The simulation also creates a configuration file that contains the coordinates of all beads in the system, printed out every 250000 cycles. By the end of the simulation, a few hundred separate system configurations are printed to this file. Using the energy and coordinate data, other system properties can be analyzed post-simulation, including the structure factor, local density distribution, percolation fraction, and heat capacity. 

\paragraph{Energy} The total energy of the system, $E$, is calculated every 1000 cycles. This is done by cycling through all beads in the system and summing the energies of all unique nearest-neighbor interactions that are found. Energies are expressed relative to the interaction energy between charges (in units of $\epsilon$).

\paragraph{Structure factor} From the system configurations generated by the simulation, the quantity \textit{S(q)}, called the \textit{structure factor}, can be calculated. The structure factor is defined in the following way~\cite{chandler}, where $\textbf{r}_j$ is the position of the \textit{j}th bead, and $N$ is the number of beads in the system:
\begin{equation}
    S(q) = N^{-1} \langle \sum_{l,j=1}^{N} \exp[i\textbf{q}\cdot (\textbf{r}_l - \textbf{r}_j)] \rangle.
\end{equation}

Here, $\textbf{q} = (n_x, n_y, n_z) 2\pi/L$ is a reciprocal lattice vector, where $n_x$, $n_y$ and $n_z$ are integers and $L$ is the system's box length, and $\left< \dots \right>$ denotes an average over all vectors with magnitude $q$ and over sampled configurations. For our purposes, we are interested in the value of $S(q)$ as $q$ approaches zero. This low $q$ limit of $S(q)$ is proportional to the compressibility of a system: a highly compressible system will have a large $S(q_{low})$~\cite{hansen}. Put differently, the structure factor of a system that contains large voids, or a system that is phase separated, rapidly increases as $q$ becomes small. This fact can be used as a measure of the heterogeneity of a system, and can help pinpoint the emergence of liquid-gas phase separation~\cite{tartaglia2005}.

\paragraph{Local density}
Analyzing local density distributions provides a method of comparing the density of different structures in a simulation~\cite{chan}. Local densities are calculated by splitting a simulation box into smaller boxes and measuring the number of occupied lattice sites in these small boxes. For each small box, a density is calculated as the number of occupied sites divided by the total number of sites in that box. These local densities are then plotted as a distribution, showing which local densities are most common for the system.

\paragraph{Percolation Fraction}

Through the analysis of system configuration data, chains can split into clusters. We define a cluster as chains connected by at least one nearest-neighbor contact between opposite charges (i.e.~chains that are electrostatically bonded). With clusters identified, system configurations can then be checked for the presence of a percolating cluster; in other words, a cluster that encounters its own periodic image, wrapping around the entire simulation box. \textit{Percolation fraction} is defined as the fraction of configurations printed throughout a particular simulation that contain a percolating cluster.

\paragraph{Heat capacity} Energy data can be used to determine the heat capacity of a system. For a constant volume system, heat capacity
is given by~\cite{chandler}
\begin{equation}
    C_v = \frac{dE}{dT} =\frac{\langle (\Delta E)^2 \rangle}{k_B T^2},
\end{equation}
where $\langle (\Delta E)^2 \rangle$ 
is the variance of the energy,
and $k_B$ is the Boltzmann constant, which we set to unity.
This expression gives two ways heat capacity can be measured from the simulations: from the slope of an energy vs.~temperature plot, or from the energy fluctuations in a simulation. Both methods are used in our analysis and produce consistent results.
 
\subsection{Sequences \& state points}
We focus our analysis on six representative sequences of 24 beads. The sequences are as follows, where n is a neutral bead, p is a positively charged bead, and g is a negatively charged bead (the number of charges in each sequence is given in brackets):
    \par \textbf{C2N0}:   \verb| nnppggppggppggppggppggnn| (20)
	\setlength{\parskip}{0pt}
	\par \textbf{C2N1}:   \verb| nppnggnppnggnppnggnppngg| (16)
	\par \textbf{C2N2}:   \verb| nppnnggnnppnnggnnppnnggn| (12)
	\par \textbf{C2N3}:   \verb| nppnnnggnnnppnnnggnnnppn| (10)
	\par \textbf{C2N4}:   \verb| nnppnnnnggnnnnppnnnnggnn| (8)
	\par \textbf{C2N16}: \verb|nnppnnnnnnnnnnnnnnnnggnn| (4)
\setlength{\parskip}{6pt} 

\noindent These sequences contain pairs of charges (C2) separated by differing amounts of neutral beads (N0 - N16). The choice of these sequences is partially motivated by the past findings suggesting that blocky charge patterns promote phase separation~\cite{fung}. For each of these sequences, a range of temperatures and packing fractions are simulated. In terms of packing fraction, a range between 0.01 and 0.46 is simulated. We simulate systems with 300 chains; the packing fraction is varied by changing the size of the cubic box. Packing fraction, denoted by $\Phi$, can be calculated by dividing the number of beads in the system ($24\times300 = 7200$) by the number of lattice sites in the box. For a box with side length $L=42$, the packing fraction is 0.097. The smallest box size that can be used for 24-length chains is $L=25$; any smaller and the chains could run into their periodic image, which is nonphysical. This puts a maximum on the packing fraction of 0.46 for 300 chains. In total, 18 to 25 temperatures at 30 different packing fractions are studied for most sequences, giving a total of over 500 simulations per sequence. Each simulation runs for roughly 3 to 4 days. 

\section{Results}

\begin{figure}[hbtp]
    \centering
    \includegraphics[width=\linewidth]{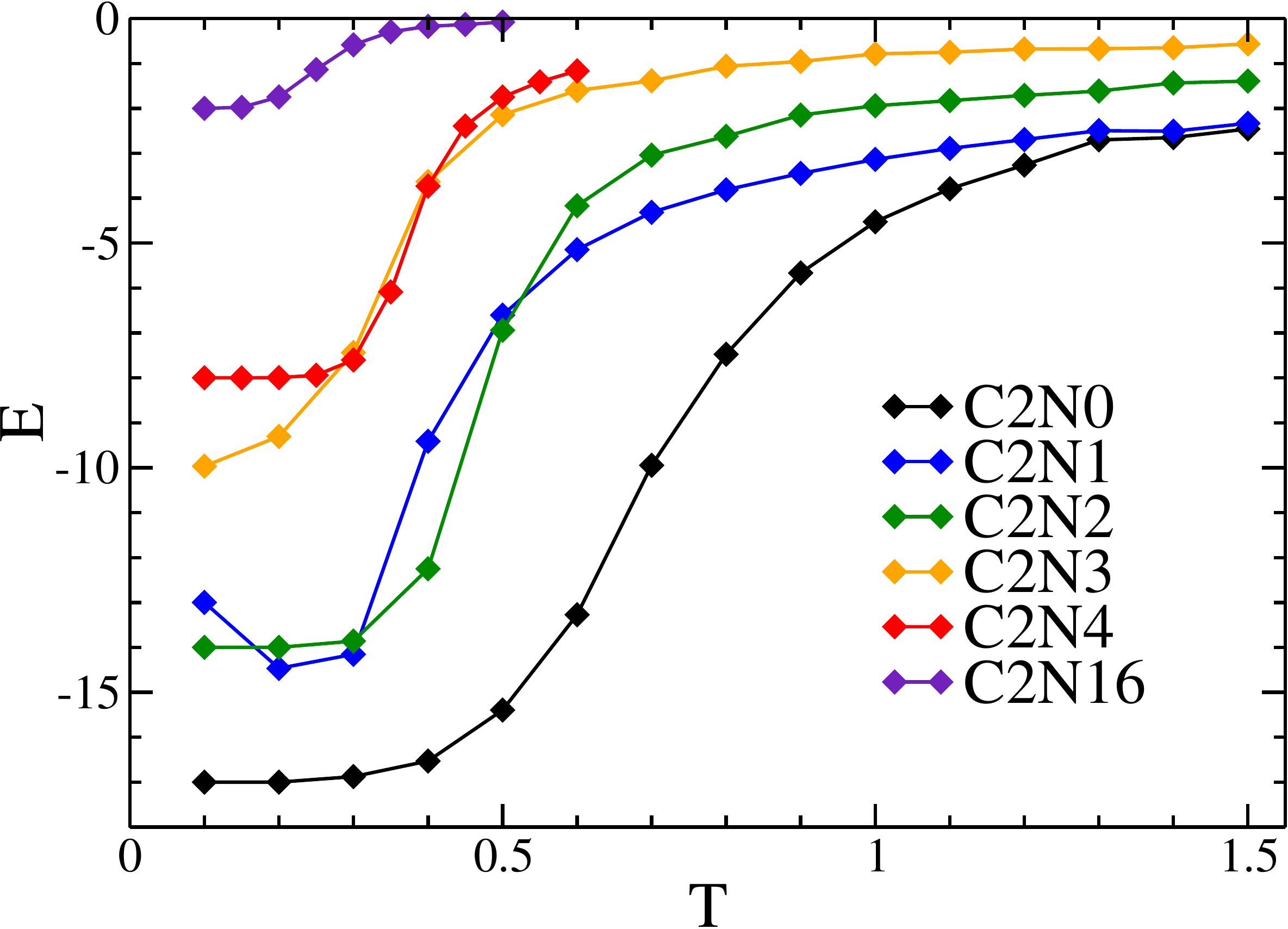}
    \caption{Average energy versus temperature for single-chain simulations with the six sequences. Each point represents a separate simulation. Total energy was averaged over the last half of each simulation to help ensure that averaging is done over equilibrated data.}
    \label{fig:onechain}
\end{figure}

Systems of 300 chains are simulated over a range of packing fraction and temperature for six different sequences. To help determine what temperature range to study for each sequence, single chain simulations are performed. Average energies of single-chain simulations at various temperatures for the six sequences are shown in Fig.~\ref{fig:onechain}. In these small-scale simulations, the temperature range at which the chain collapses in on itself and forms a low-energy structure roughly corresponds to the temperature range where interesting structural transitions occur for a bulk system of chains with that sequence. Sequences with more charges tend to collapse, condense or crystallize to low energy structures at higher temperatures. 

\begin{figure*}[hbtp]
	\includegraphics[width=\linewidth]{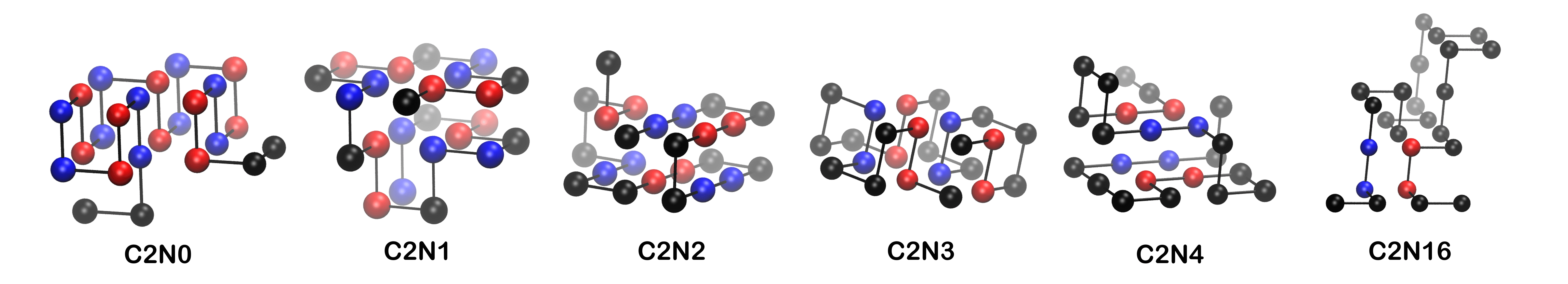}
	\caption{\label{fig:minch}Minimum energy structures for the six sequences, identified from low temperature Monte Carlo simulations with one chain. These conformations are not necessarily unique. In these snapshots, positive charges are shown in red, negative charges in blue, and neutrals in black. All images in this work are created using Visual Molecular Dynamics (VMD)~\cite{vmd} and rendered using Tachyon~\cite{tachyon}.}
\end{figure*}

\begin{figure*}[hbtp]
	\includegraphics[width=0.9\linewidth]{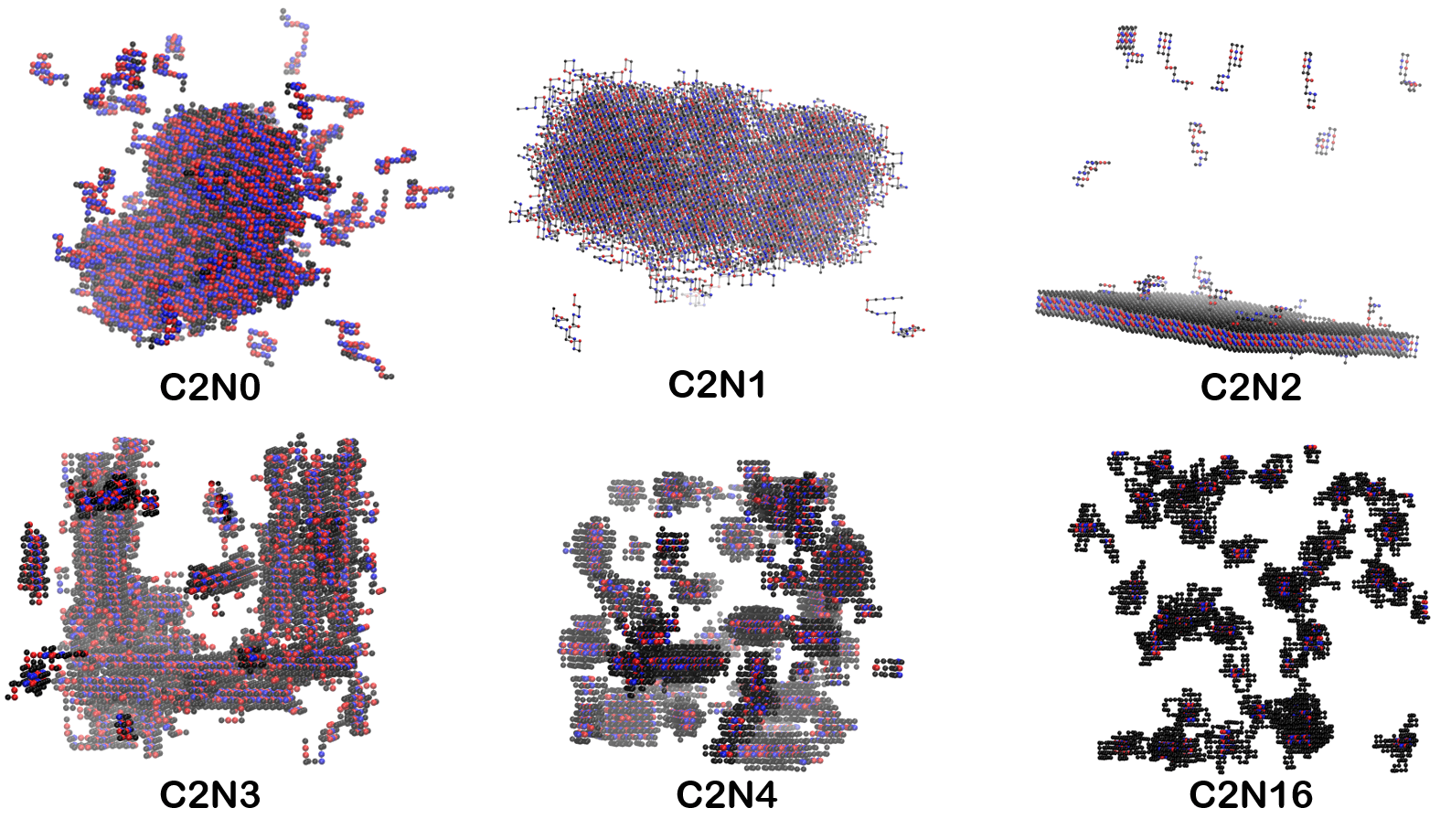}
	\caption{\label{fig:dim}The rich variety of low temperature behavior for these six sequences. As the length of the neutral patches along the sequence is increased, the dimensionality of the low-energy, low-temperature structure decreases. The state points depicted in these snapshots are as follows: \textbf{C2N0}: $T$=1.025, $\Phi$=0.097. \textbf{C2N1}: $T$=0.575, $\Phi$=0.097. \textbf{C2N2}: $T$=0.55, $\Phi$=0.097. \textbf{C2N3}: $T$=0.34, $\Phi$=0.097. \textbf{C2N4}: $T$=0.35, $\Phi$=0.097. \textbf{C2N16}: $T$=0.25, $\Phi$=0.097. Positive charges are shown in red, negative charges in blue, and neutrals in black.}
\end{figure*}

Minimum energy structures for each sequence are shown in Fig.~\ref{fig:minch}. These conformations are not necessarily unique -- neutral patches along the sequences are especially flexible -- but they demonstrate how charge pairs can arrange to optimize the energy of the chain. In some cases, these single-chain structures provide insight into the types of bulk structures that form in large systems of these sequences. The C2N0, C2N2 and C2N3 single-chain conformations, for instance, can act as building blocks in forming the bulk structures seen in low temperature systems of these sequences (shown in Fig.~\ref{fig:dim}). 


C2N1 has a much more complex single-chain minimum energy structure. A negative slope in $E(T)$ is visible at low temperature for this sequence (Fig.~\ref{fig:onechain}); at the lowest temperature simulated, $T=0.1$, the chain does not find this true minimum energy structure, even after extending the simulation to over $10^{10}$ Monte Carlo steps. The chain instead becomes frozen in a slightly higher energy structure that is much less complex and very conformationally distant from the optimal structure. The degree of structural fluctuation present at this temperature is simply too low to reach the true minimum energy structure in a reasonable amount of simulation time. Compared to other sequences, a C2N1 chain has greater difficulty folding into an optimal structure on its own.

The low-temperature structures formed in systems of 300 chains are shown for each sequence in Fig.~\ref{fig:dim}. These structures are highly sequence dependent, varying greatly as the length of neutral patches in the sequence is changed. The C2N0 and C2N1 systems form large, roughly spherical crystal structures. C2N2 collapses into a thin sheet with charge pairs sandwiched between layers of neutral beads. C2N3 forms cylindrical stalks with charged cores and insular neutrals at their surface. C2N4 and C2N16 collapse into small, finite-sized clusters with charges sequestered at their cores. As the length of neutral patches is increased, the dimensionality of these structures decrease, from three-dimensional spheres, to two-dimensional plates, to one-dimensional filaments, to zero-dimensional, point-like clusters which do not increase in size. The behavior of these six sequences will be discussed in turn.

\subsection{C2N0 \& C2N2:\\Fluid-Crystal Transition}

\begin{figure}[htbp]
    \includegraphics[width=\linewidth]{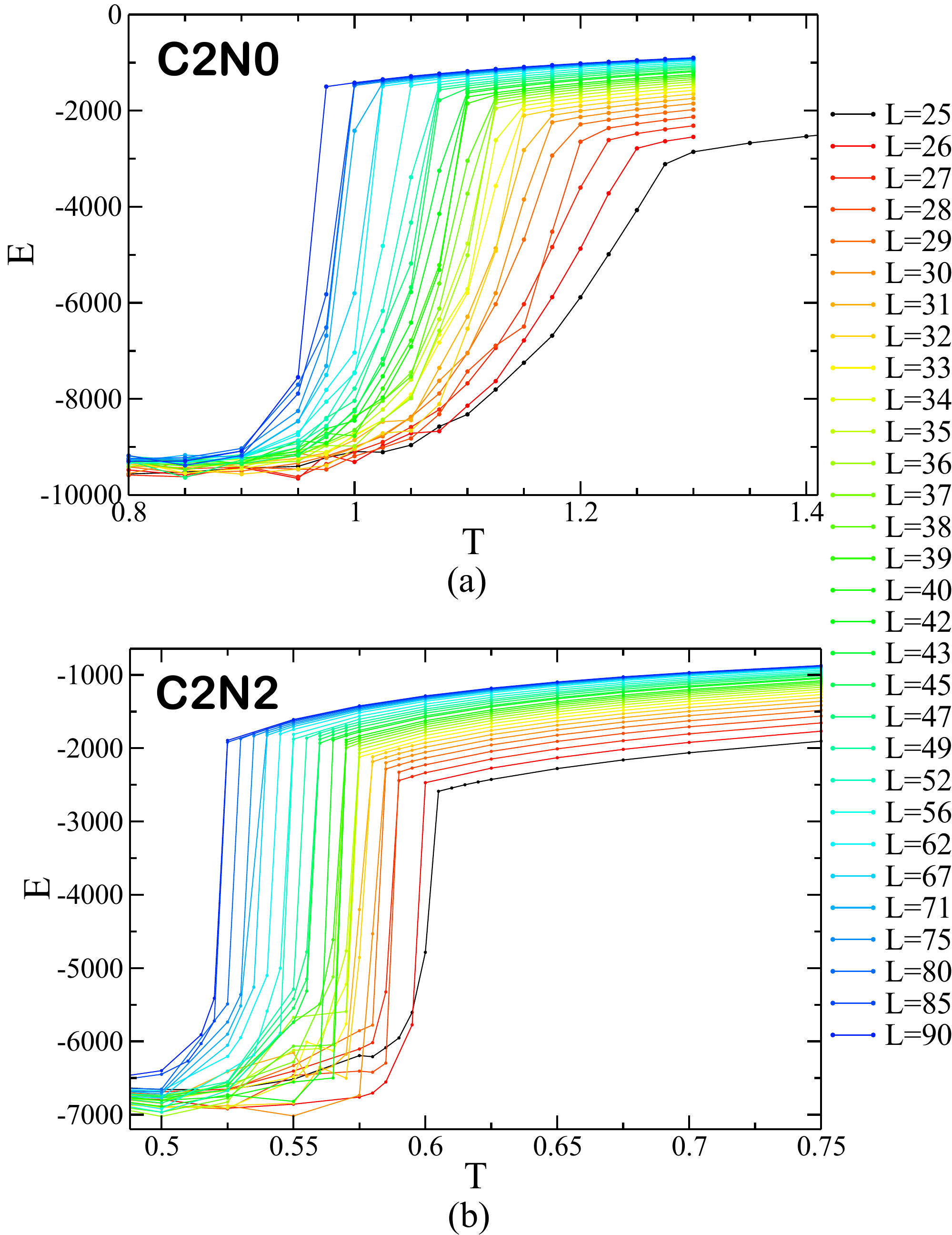}
 	\caption{\label{fig:c2n20_enr}Average energy versus temperature plots for systems of (a) 300 C2N0 chains or (b) 300 C2N2 chains. Every point represents a separate simulation. Energy is averaged over the last quarter of each simulation to help ensure that averaging is done over equilibrated data. Lines represent different packing fractions ranging from $\Phi=0.0099$ ($L=90$) to $\Phi=0.461$ ($L=25$). The legend corresponds to both figures.}
\end{figure}

For C2N0 and C2N2, simulations were conducted with 300 chains at 30 different packing fractions between $\Phi=0.0099$ and $\Phi = 0.46$. C2N0 is simulated in the temperature range of $T=0.600$ to $T=1.300$, while C2N2 is simulated in the range $T=0.400$ to $T=0.950$. The choice of temperature range is informed by the single chain behavior seen in Fig.~\ref{fig:onechain}. The average total energy for each state point for both sequences is shown in Fig.~\ref{fig:c2n20_enr}. At every simulated packing fraction, there exists a temperature where an abrupt change in energy occurs. This change in energy is associated with a phase transition. The chains condense from a fluid state into a phase separated state of rare fluid and dense crystal,
 as expected for this first-order transition occurring at fixed volume. 
Systems abruptly shift from a uniformly dispersed configuration to a dense, organized, clustered state. No liquid-like behavior is observed above these crystal transitions; once chains begin condensing together and interacting favorably, crystals rapidly form. The crystal structure is sequence-dependent: the low-energy structures formed by these two sequences are vastly different, as seen in Fig.~\ref{fig:dim}. These crystal transitions display the metastability characteristic of a first-order phase transition~\cite{debenedetti}. Metastable behavior is seen when temperatures very near the transition are simulated. Systems could persist in the high-energy fluid state for an extended period of time before crystallizing. 

\begin{figure}[htbp]
    \subfloat[]{\label{fig:c2n20_pda}%
        \includegraphics[width=0.86\linewidth]{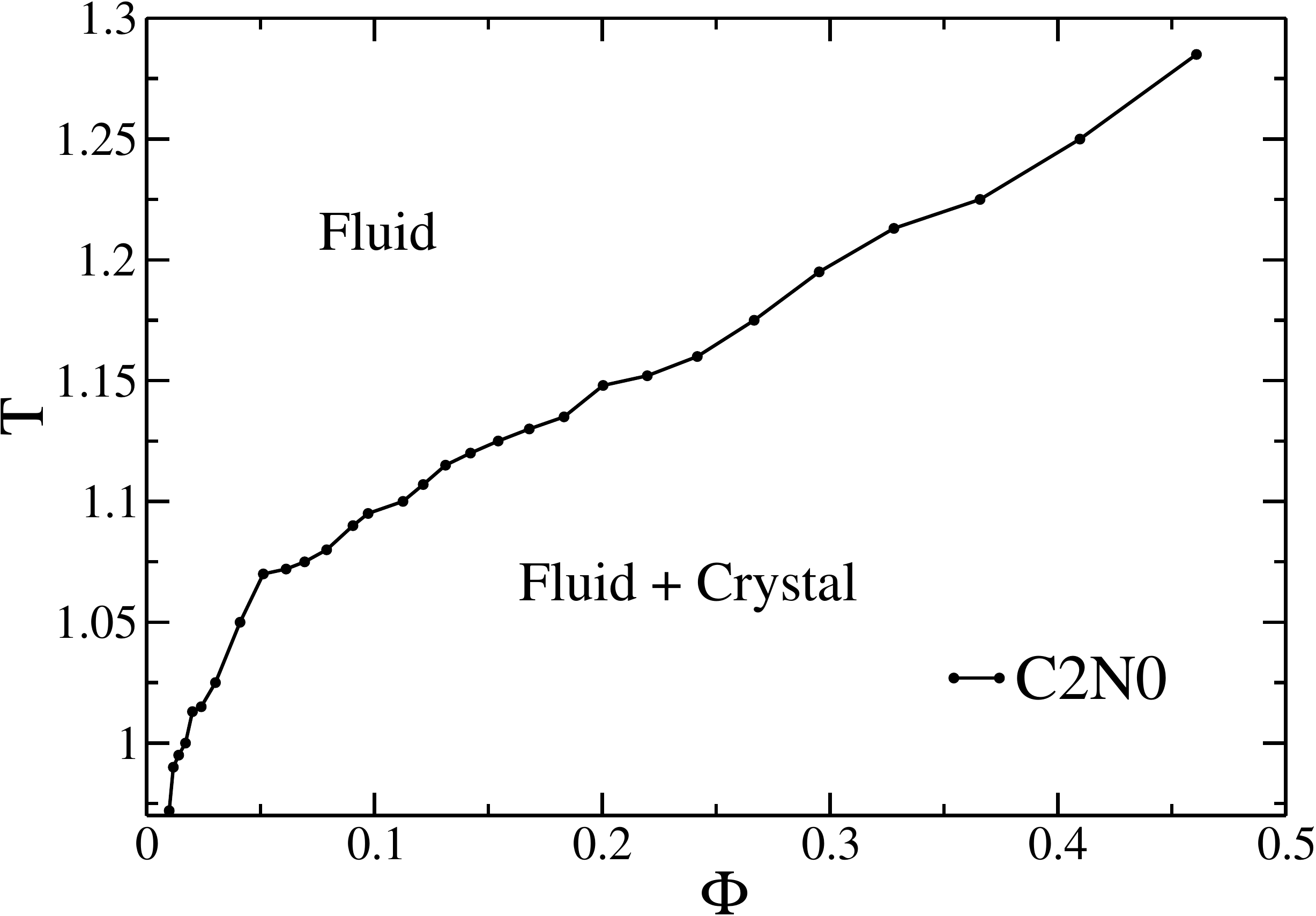}%
    }
    
    \subfloat[]{\label{fig:c2n20_pdb}%
        \includegraphics[width=0.86\linewidth]{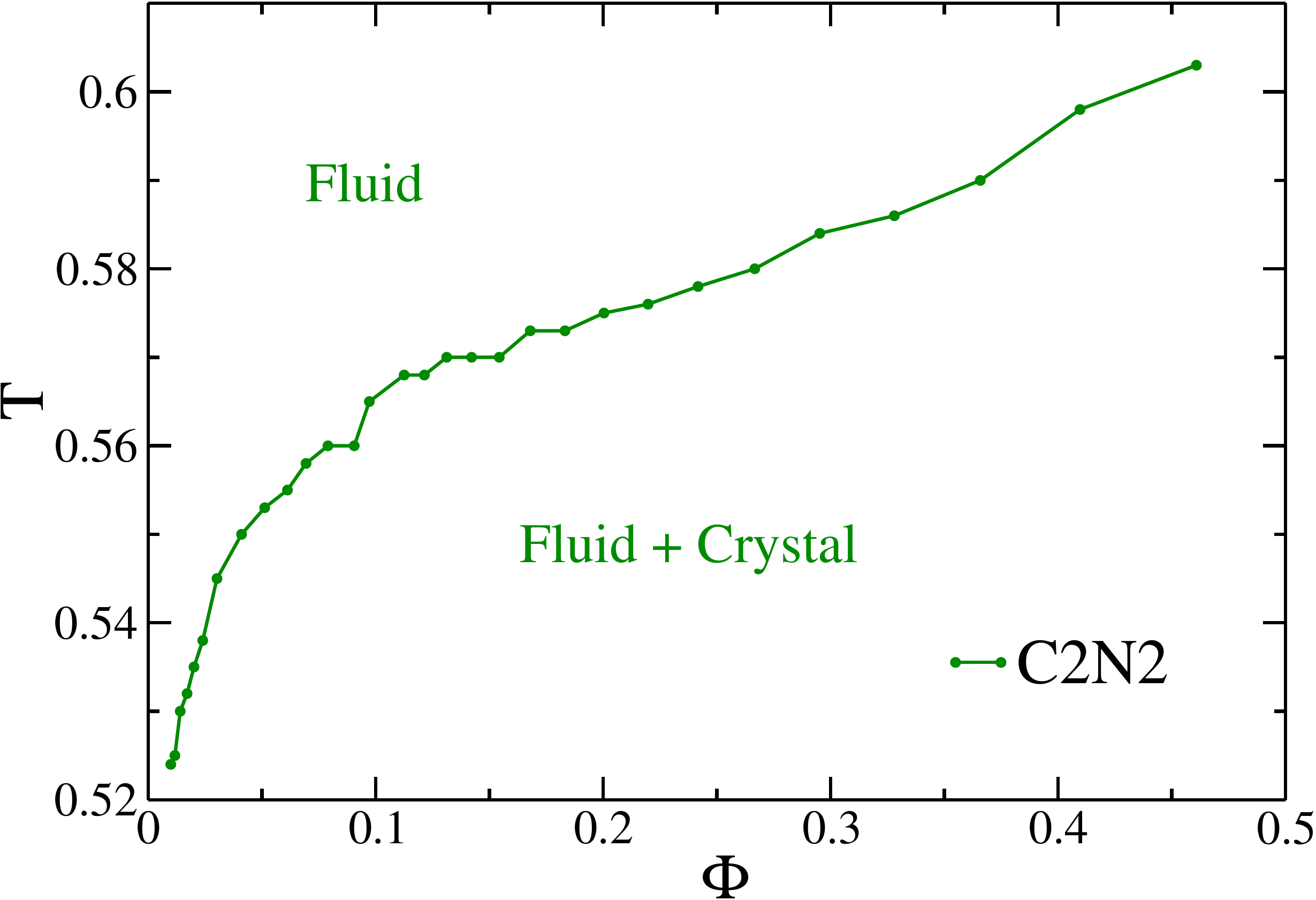}%
    }
 	\caption{\label{fig:c2n20_pd}Phase diagrams for the C2N0 and C2N2 sequences. The Fluid-crystal coexistence curve is approximated from $E(T)$ data at each packing fraction. The crystal densities are near unity; these are not shown in the diagram. $C_V$ maxima do not occur above the crystal transition, so they are not included in the figure.}
\end{figure}

Phase diagrams can be constructed for these sequences from the energy data. These are shown in Fig.~\ref{fig:c2n20_pd}. The location of the fluid to crystal transition is approximated from where, as temperature is decreased, a kink first appears in the $E(T)$ plot. This `kink' marks the formation of small, low-energy crystals. Below this temperature, crystal and fluid phases form an equilibrium. The crystal phase grows and incorporates more chains as temperature is decreased further. For higher packing fractions, crystal formation occurs at higher temperature. The shape of this crystal boundary is similar for both of these sequences, but they crystallize over very different temperature ranges. This could be explained by the difference in their charge densities: C2N2 contains 12 charges, while C2N0 contains 20 -- almost double. Similarly, C2N0 crystallizes at temperatures almost double those seen for C2N2. 

While they display similar phase behavior, the crystal structures of C2N0 and C2N2 are considerably different. The reason behind this can easily be understood by comparing their sequences. 

\begin{figure}[htbp]
    \subfloat[]{\label{fig:c2n20_closeupa}%
         \includegraphics[width=0.33\linewidth]{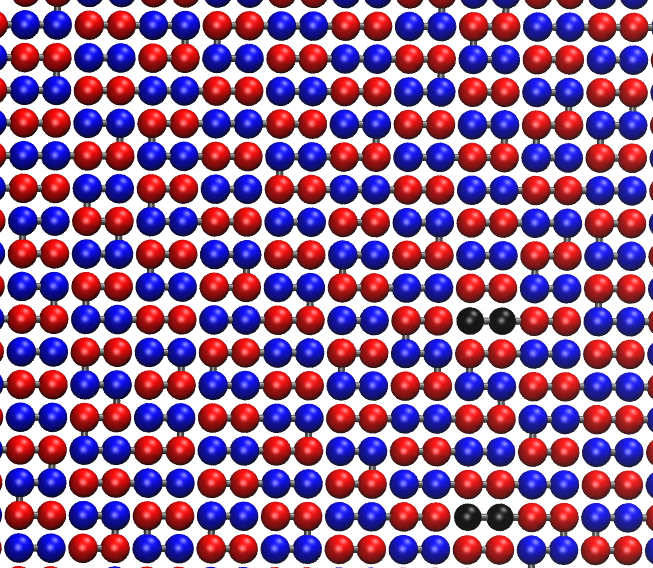}%
    }
    \subfloat[]{\label{fig:c2n20_closeupb}%
        \includegraphics[width=0.45\linewidth]{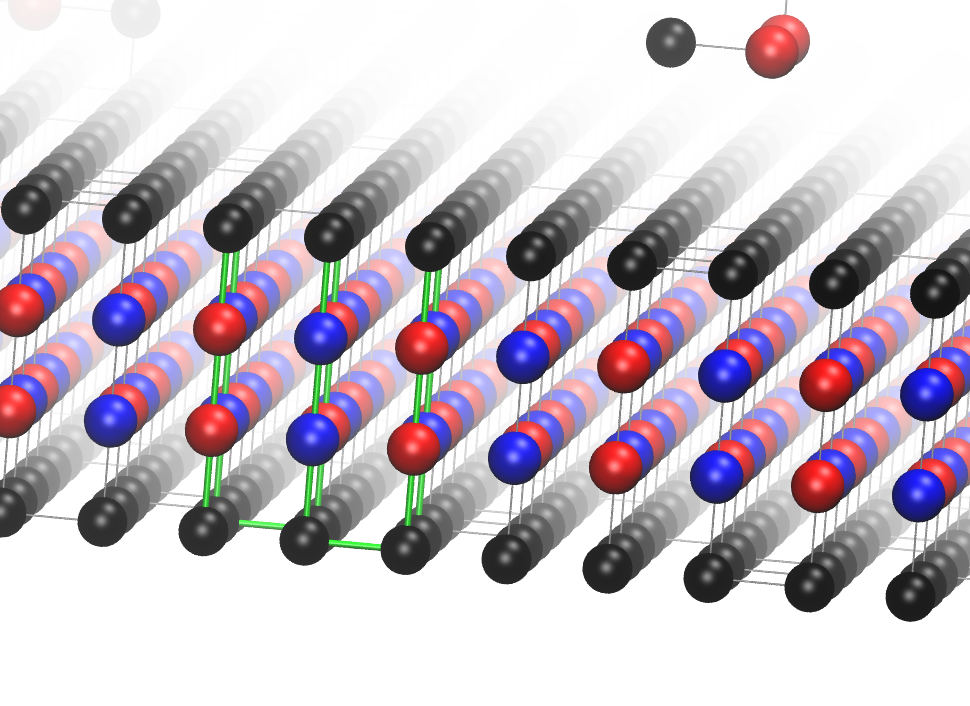}%
    }
 	\caption{\label{fig:c2n20_closeup}Closeup views of the crystal structures for C2N0 and C2N2. In (a), a single layer of a C2N0 crystal is shown at $T=1.025$ and $L=42$ ($\Phi=0.097$). Charges are highly organized. Most neutral beads are sequestered to the surface of the crystal. In (b), a side-on view of the highly organized, membrane-shaped structure of C2N2 chains is shown at $T=0.55$ and $L=42$. Positive charges are shown in red, negative charges in blue, and neutrals in black. In (b), a single chain is highlighted in green.}
\end{figure}

C2N0 consists of five pairs of positive charges and five pairs of negative charges arranged in an alternating sequence, with two neutral beads on each end (nnppggppggppggppggppggnn). C2N0 chains could therefore simply stack together in an anti-parallel fashion to create a large, three-dimensional cluster where all charges form optimal electrostatic interactions. An interior view of a layer in a C2N0 crystal is shown in Fig.~\ref{fig:c2n20_closeupa}. Charges form a checkerboard pattern, with most neutrals exposed to the outside of the crystal. The pattern of charges is highly organized, but the conformations of individual chains are fairly heterogeneous; chains are not organized in a specific way but are instead folded into many different configurations within the crystal, while still maintaining the organized charge pattern. 

C2N2 contains two neutral beads separating each pair of positive charges and pair of negative charges (nppnnggnnppnnggnnppnnggn). On a lattice, those two neutral beads separating charges allow the chain to fold back on itself and form favorable interactions between charge pairs. On a larger scale, this results in the crystal shown in Fig.~\ref{fig:c2n20_closeupb}. Each chain can fold into a separate building block which combine to form the organized structure of the crystal. There is less conformational heterogeneity in the way chains are organized in this structure. This two-dimensional, membrane-like cluster extends through a slice of the simulation box, eventually encountering its periodic image. At higher packing fractions, several of these flat crystals form, sometimes stacking parallel or running perpendicular to each other. For both C2N0 and C2N2, chains can individually fold into energetically favorable configurations which can then combine to give the crystal state.

\begin{figure}[htb]
	\includegraphics[width=0.92\linewidth]{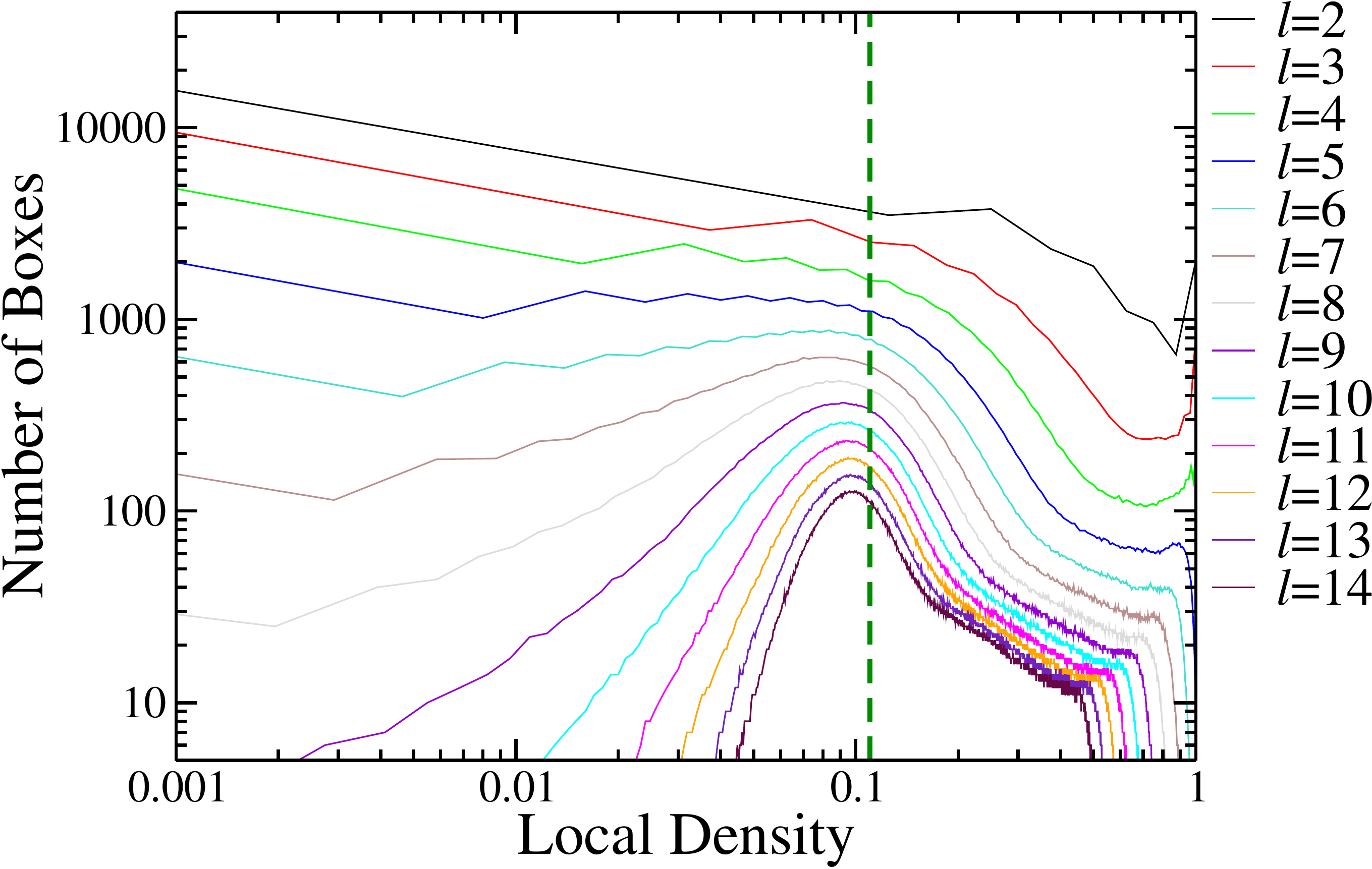}
	\caption{\label{fig:c2n0_ld}C2N0; $T=1.1$; $L=34$; $\Phi=0.183$. Distribution of local density for a state point below the C2N0 crystal transition temperature. The curves indicate the side lengths of different small boxes used in the calculation. The simulation box is divided into $34^3=39304$ small boxes of side length $l$ (one centered at each lattice site). The distribution shows how many of these small boxes have a particular density. The coexisting fluid density measured from the phase diagram is plotted as a dashed green line.}
\end{figure}

A local density distribution for a C2N0 system is shown in Fig.~\ref{fig:c2n0_ld}. This is generated at a state point where a crystal has formed. For the system studied, the distribution fully extends from zero to one. This suggests that the configuration is highly heterogeneous: some parts of the simulation box have very low density, while others are completely condensed. For smaller local box sizes, a peak in the local density distribution occurs at 1. This corresponds to the interior of the crystal. 

This phase separated system can be considered in terms of its coexistence densities. In this case, the crystal has a density of 1.
From the phase diagram plotted in Fig.~\ref{fig:c2n20_pda}, the coexistence density of the fluid at $T=1.1$ is approximately $p_f\approx0.11$. This value is plotted on Fig.~\ref{fig:c2n0_ld}. Near this value, there is a peak in the local density distribution. Densities above this value become less common, until eventually reaching a peak again near 1, which corresponds to the crystal interior. This local density behavior supports that the measurement of fluid coexistence density from the phase diagram is reasonable for this system. Analyzing local density distributions present another method of constructing phase diagrams that will be explored in the following section. Local maxima in the distribution can approximate the coexistence densities of phases. 

\subsection{C2N1:\\Liquid Condensation}

The C2N1 sequence, which contains single neutral beads separating pairs of charges, exhibits three distinct types of structure: a homogeneous fluid phase, a liquid-like droplet, and a dense crystal. These structures appear at specific ranges of temperature and packing fraction. The phase behavior of this sequence can be characterized through analysis of energy, structure factor, local density distributions, and specific heat data.

\begin{figure*}[hbtp]
	\includegraphics[width=0.92\linewidth]{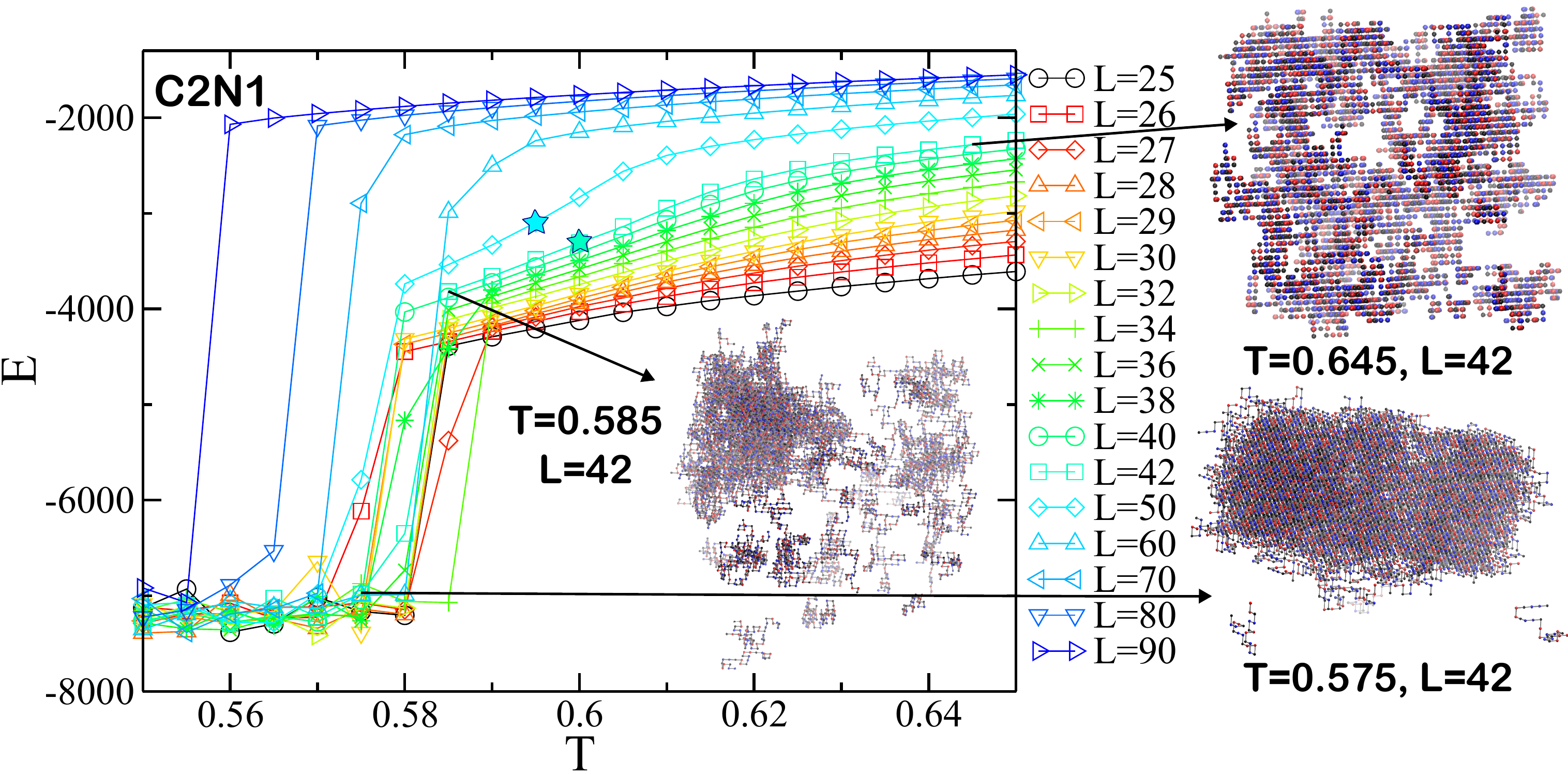}
 	\caption{\label{fig:c2n1_enr}Energy versus temperature plot for systems of 300 C2N1 chains with various packing fractions (denoted by box lengths L). Every point represents a separate simulation. Total energy is averaged over the last quarter of each simulation to help ensure that averaging is done over equilibrated data. For L=42 and L=50, the inflections in $E(T)$ which mark the formation of a liquid droplet are highlighted by stars. Images from three state points are shown for $L=42$ ($\Phi=0.097$). In these images, positively charged beads are shown in red, negative beads in blue, and neutral beads in black.}
\end{figure*}

Systems of 300 C2N1 chains are simulated at packing fractions ranging from $\Phi=0.0099$ to $\Phi = 0.46$. Temperatures range from 0.55 to 0.65, with a resolution of 0.005 between simulated state points. We simulate some packing fractions at temperatures outside of this range, but no other interesting behavior is observed. The average system energy at each state point is plotted in Fig.~\ref{fig:c2n1_enr}. 

Like C2N0 and C2N2, the C2N1 sequence exhibits an abrupt fluid to crystal transition. However, C2N1 systems display liquid-like behavior above the transition to a crystal. The chains collapse into a highly deformable and mobile liquid droplet. Here, chains begin to form bonds through favorable electrostatic interactions but are yet fully organized into a crystal state. Droplets stably form and endure throughout the length of simulations. For these systems of 300 chains, the onset of a liquid state is characterized by an inflection in $E(T)$ rather than a clear break in slope like what is seen for the crystal transition.  
Since our simulations are at constant volume, $E(T)$ should exhibit a kink as the system phase separates into a protein-rich phase (liquid) and a protein-poor phase (gas).  As we show below, we recover this behaviour for larger system size.
These droplets form gradually over a range of temperatures, with droplets becoming denser closer to the crystal transition temperature.
The system contains small, local heterogeneities at higher temperature; these grow more prominent as temperature is decreased, eventually resulting, at low enough packing fraction, in the formation of a liquid droplet with large voids in the rest of the system. 
%%%IVAN
This is characteristic of condensation, a first-order transition, occurring at constant volume.  Even for simulations carried out along the critical isochore, where phase separation would occur continuously via a second-order transition, similar observations would be made.
At higher packing fractions, a continuous, system-spanning, liquid-like network of chains forms. At even lower temperatures, chains condense further into a crystal.

\begin{figure}[htb]
	\includegraphics[width=0.8\linewidth]{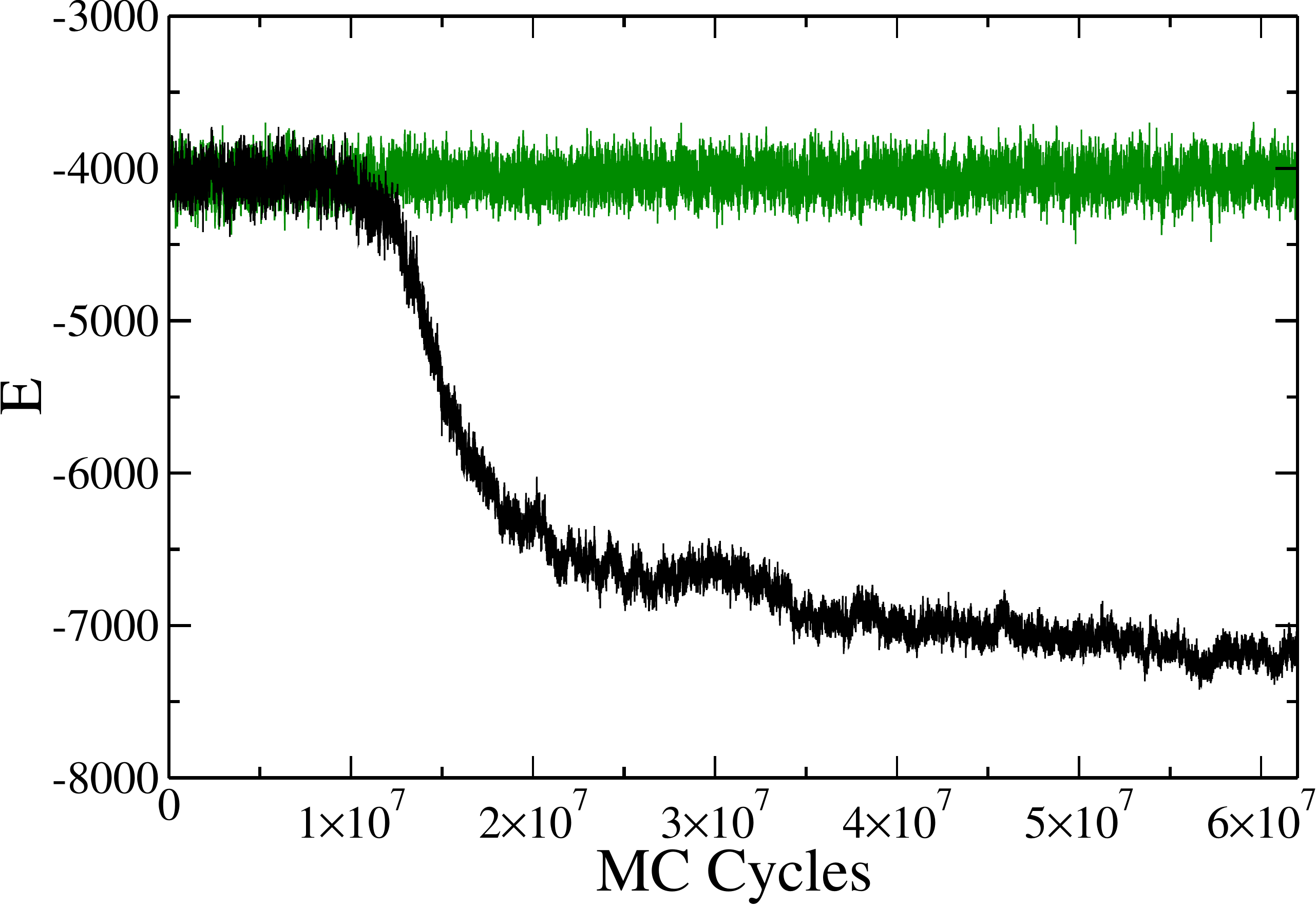}
 	\caption{\label{fig:meta_c2n1}The energy progression of two separate C2N1 simulations conducted at the same state point, $T=0.585$ with a box length of $L=34$ ($\Phi=0.183$). The energy is sampled every 1000 Monte Carlo cycles, with 7200 steps per cycle. In one simulation, the system remains in a higher energy state -- a crystal never forms. In the other simulation, the energy of the system begins decreasing after 10 million cycles. This decrease in energy is associated with the formation of a crystal.}
\end{figure}

There is a high degree of metastability near the crystal transition. State points around the transition temperature can stably exist in either the high-energy (liquid/gas or fluid) state, or the low-energy crystal state. This is illustrated in Fig.~\ref{fig:meta_c2n1}. The energy of the crystallized system, seen in the latter half of the black line in the figure, equilibrates very slowly; this is also the case for the crystal states of C2N0 and C2N2, as well as the low-temperature collapsed states of C2N3, C2N4 and C2N16. On the other hand, the energy of C2N1's liquid state, as shown by the green line in Fig.~\ref{fig:meta_c2n1}, appears to quickly reach a stable equilibrium in simulations. In some simulations conducted very near this transition temperature, over a hundred billion steps are required for a crystal to begin forming. Crystal growth is instigated by the formation of a small, low-energy, ordered structure -- the nucleus of the crystal. Chains must fold into a specific, ordered conformation in order to promote the propagation of the solid structure. Once a crystal begins to form, it rapidly incorporates more chains and the system energy drops precipitously. From the simulations, nucleation appears to occur over a range of temperatures but is increasingly rare at higher temperatures. Metastable behavior is observed over a temperature range of approximately $\Delta T=0.01$.
This metastability makes it difficult to pinpoint the transition temperature of the crystal state.

\begin{figure}[htbp]
    \subfloat[]{\label{fig:closeup_c2n1a}%
    	\includegraphics[width=0.7\linewidth]{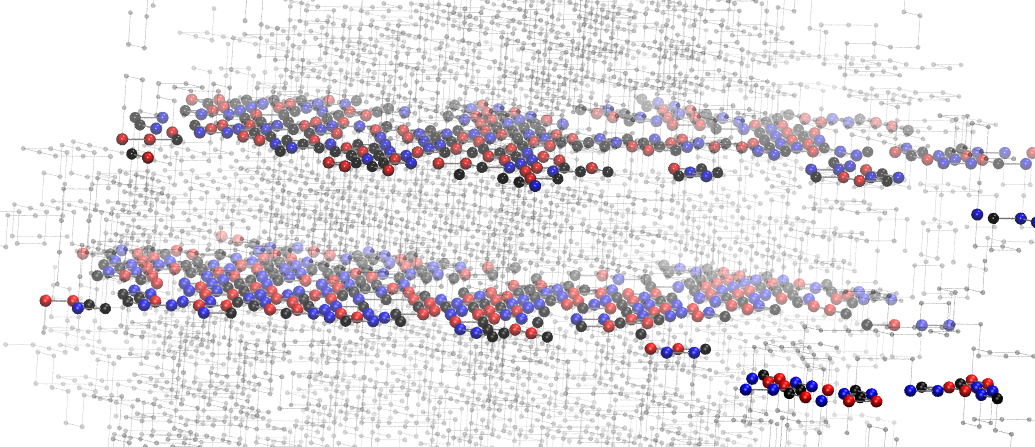}%
    }
    
    \subfloat[]{\label{fig:closeup_c2n1b}%
    	\includegraphics[width=0.7\linewidth]{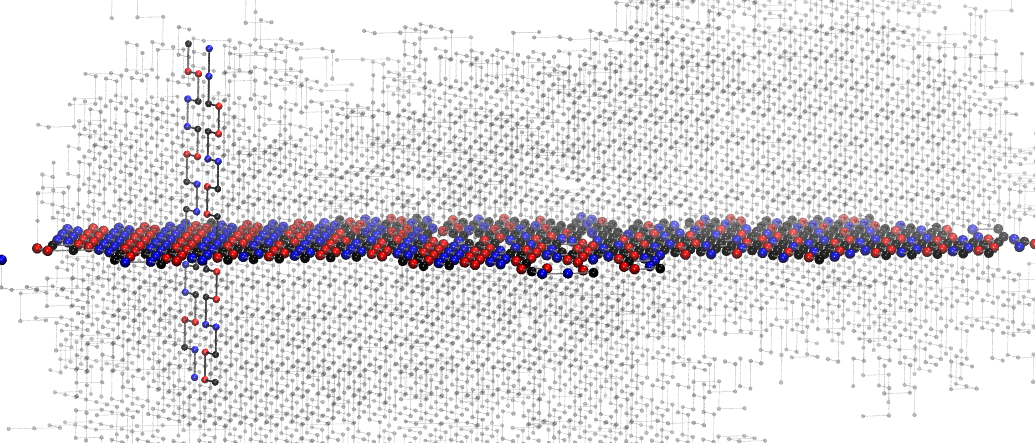}%
    }
 	\caption{\label{fig:closeup_c2n1}Closeup image of a droplet and crystal for C2N1 at $\Phi=0.097$. Positive charges are shown in red, negative charges in blue, and neutrals in black. In (a), an image of a liquid droplet is shown at $T=0.585$ and $L=42$ ($\Phi=0.097$). Two slices along x=6 and x=11 illustrate the internal structure of the droplet. The droplet is not fully condensed and displays little organization. A crystal is shown in (b) at $T=0.575$ and $L=42$. The slice along x=20 is highlighted, along with two adjacent chains. In this structure, the charges form very organized interactions.}
\end{figure}

Structural differences between the droplet and crystal phases are highlighted in Fig.~\ref{fig:closeup_c2n1}, showing closeup snapshots from the same state points as the images in Fig.~\ref{fig:c2n1_enr}. The droplet is not fully condensed; there are many empty lattice sites within its structure. Charges in the liquid-like cluster tend to form favorable interactions between chains, but there is little overall organization to the structure. The crystal, on the other hand, is very organized. Pairs of charges are arranged in alternating rows. The neutral beads along the C2N1 sequence create some disorder within the crystal. 

The C2N1 sequence consists of one neutral bead positioned between every pair of charges. Because of these lone neutrals, a single C2N1 chain cannot readily fold back on itself into an optimal, low-energy configuration (as demonstrated in Fig.~\ref{fig:onechain} and Fig.~\ref{fig:minch}). To form optimal charge interactions and build a crystal, multiple chains must agglomerate together into a complex inter-chain assembly, exemplified by the two chains highlighted in Fig.~\ref{fig:closeup_c2n1}b. This could explain why liquid droplets form above the crystal transition. Single C2N1 chains are disordered and energetically frustrated; they must form an intricate multi-chain configuration to maximize charge interactions. Due to its complexity, the crystal state has a decreased tendency to form; in its place, a partially condensed, liquid-like state becomes more thermodynamically favorable at intermediate temperatures. This is unlike C2N0 and C2N2, where the even nature of their sequence patterns make intra-chain folding more straightforward and energetically favorable. C2N0 or C2N2 chains readily fold into building blocks which can combine to build a crystal. This may explain why an intermediate liquid-like structure before crystal formation is not seen for C2N0 or C2N2 but is for C2N1.

\begin{figure}[htb]
    \subfloat[]{\label{fig:c2n1_sqa}%
        \includegraphics[width=0.84\linewidth]{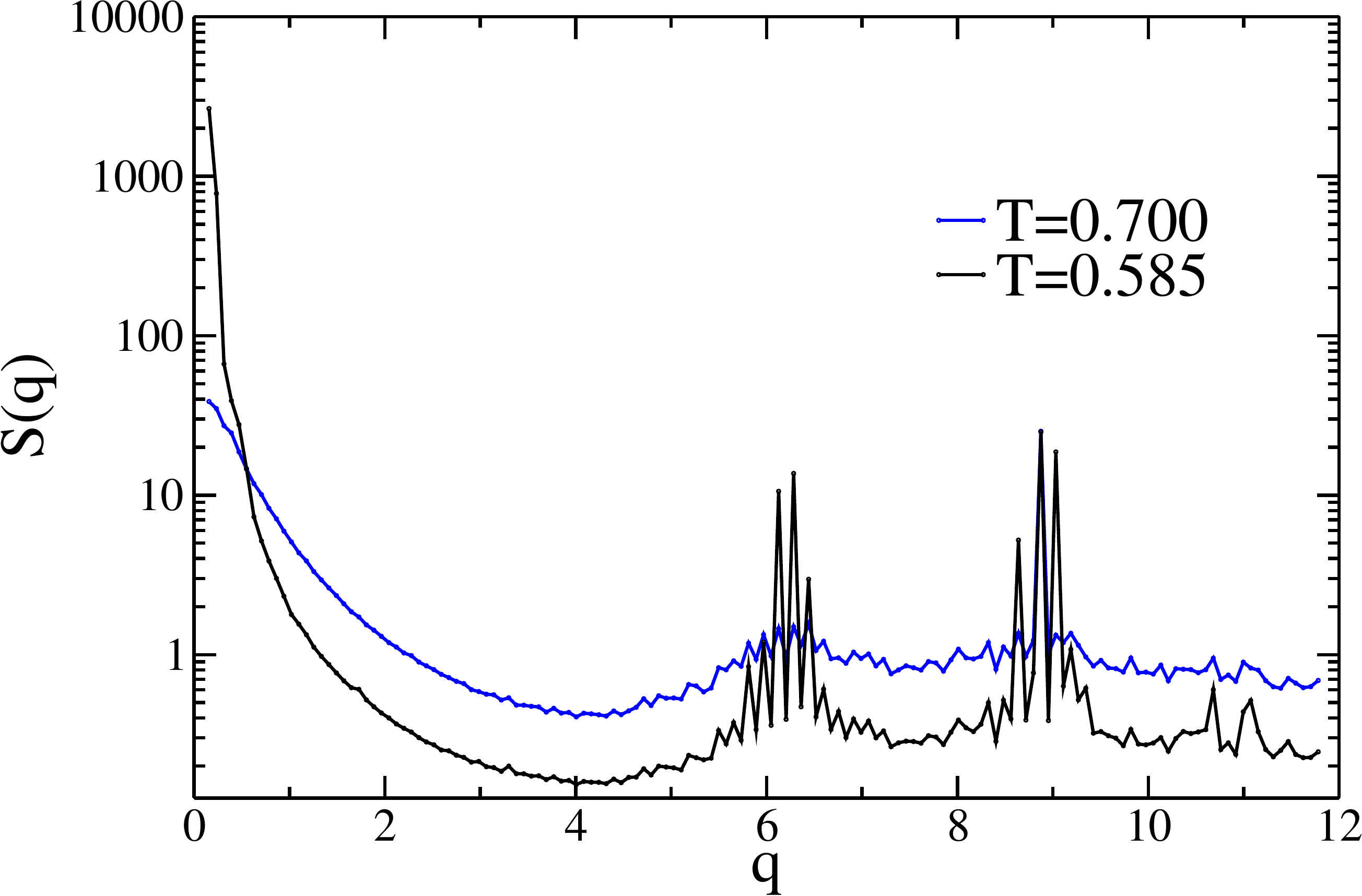}%
    }
   
   \subfloat[]{\label{fig:c2n1_sqb}%
   		\includegraphics[width=0.84\linewidth]{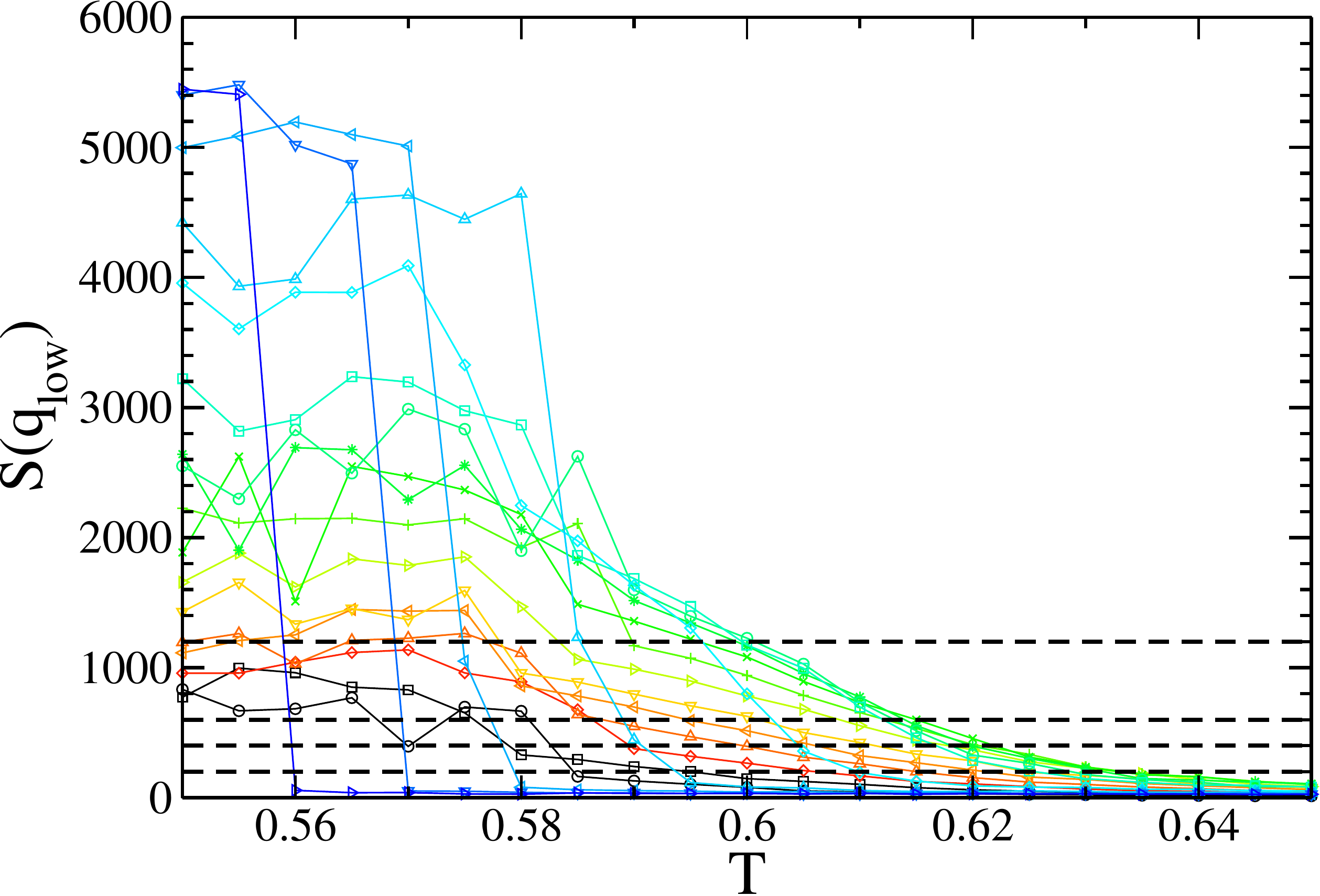}%
   }
   \caption{\label{fig:c2n1_sq}(a) Structure factor data for $\Phi=0.097$ at two different temperatures for systems of 300 C2N1 chains. (b) The values of $S(q)$ at lowest $q$ for each state point plotted versus temperature. Colored lines correspond to different packing fractions (colors have the same meaning as in Fig.~\ref{fig:c2n1_enr}). The black dotted lines mark four chosen thresholds in $S(q_{low})$ (200, 400, 600, and 1200).}
\end{figure}

The structure factor is a useful quantity for characterizing the behavior of these systems. The low $q$ behavior of $S(q)$ gives information about the heterogeneity and compressibility of a system, which can be useful for identifying the formation of a liquid droplet. Some sample structure factor data for C2N1 are shown in Fig.~\ref{fig:c2n1_sqa}. At higher temperature, when chains are uniformly dispersed throughout the simulation box, the maximum in structure factor at low $q$ is below 100. At lower temperature, in a system where chains have condensed into a liquid-like droplet, the structure factor reaches above 2000 at the lowest $q$ ($q_{low}=2\pi/L$). When chains condense into a droplet, empty voids are left in the rest of the simulation box. Structure factor detects this heterogeneity in the system, which is then seen as an elevated peak at $S(q_{low})$. The peaks in the structure factor that occur at higher values of $q$ on Fig.~\ref{fig:c2n1_sqa} are due to the discrete nature of the lattice.

\begin{figure}[htbp!]
    \includegraphics[width=\linewidth]{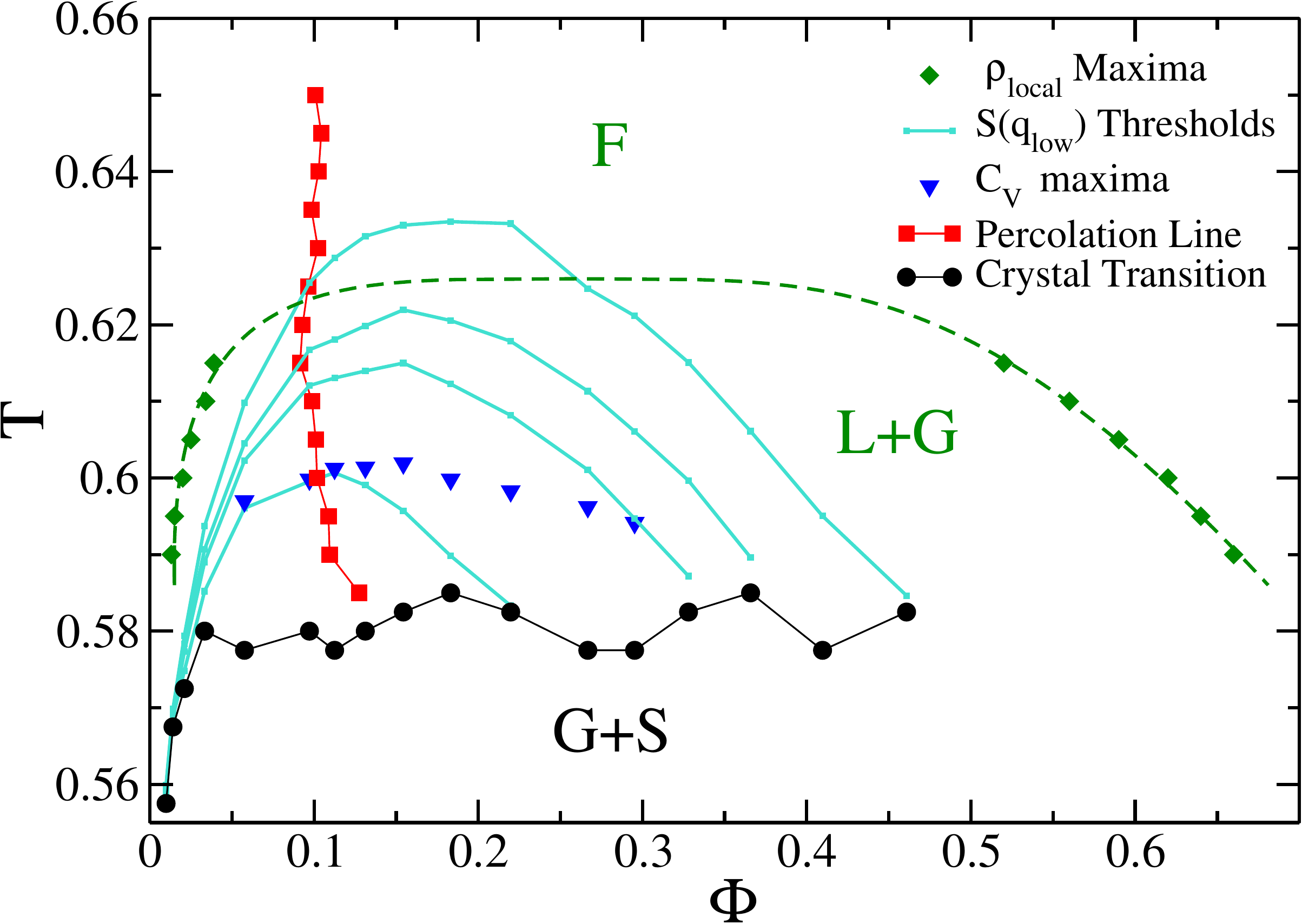}
 	\caption{\label{fig:pd_c2n1}A phase diagram for the C2N1 sequence. F=fluid, G=gas, L=liquid, and S=solid. The location of the peaks in local density at each temperature for the 2400 chain systems are shown in green, along with a fit produced from this data (dashed line). The rest of the information plotted is derived from 300 chain simulations. Points on the turquoise lines correspond to the temperatures that crossed a chosen threshold in $S(q_{low})$. These temperatures were linearly interpolated from the $S(q_{low})$ data shown in Fig.~\ref{fig:c2n1_sqb}. Four different thresholds are plotted: $q_{low}=$ 200 (topmost curve), 400, 600 and 1200 (bottommost curve). The crystal boundary, shown in black, is determined from the abrupt drop in $E(T)$ (Fig.~\ref{fig:c2n1_enr}). The heat capacity maximum for each packing fraction is shown in blue. Only local maxima that occurred above the crystal transition were considered. The percolation line for this system is also plotted (red); packing fractions to the right of this line were percolated more than 50\% of the time during simulations.}
\end{figure}

With this in mind, structure factor measurements can be used to classify the state points where liquid droplet formation occurs, as shown in Fig.~\ref{fig:c2n1_sqb}. By looking at lowest-$q$ values of the structure factor, thresholds in $S(q_{low})$ can identify trends in the compressibility and heterogeneity of systems, thus helping to elucidate phase behavior~\cite{tartaglia2005}. One threshold is chosen such that the maxima in heat capacity, plotted in Fig.~\ref{fig:pd_c2n1}, aligns with the temperatures that cross the threshold. These heat capacity maxima are calculated from the slope in $E(T)$. When liquid droplets begin forming, the system shifts from a homogeneous fluid configuration to a configuration where chains favorably interact to form a dynamic cluster; this roughly coincides with a maximum in energy fluctuations and a maximal slope in $E(T)$, i.e.~a peak in the heat capacity. Systems with the highest packing fractions never cross this threshold in $S(q_{low})$; for these systems, a well-defined liquid droplet does not form before the crystal transition, although the systems do exhibit liquid-like behavior. These denser systems remain mostly homogeneous, only containing relatively small voids, and with chains forming an increasingly favorable network of interactions as temperature is lowered. Liquid droplets do not form through an abrupt transition, instead gradually condensing over a range of temperatures. Therefore, this choice of threshold for $S(q)$ is not unique. Four different threshold values are plotted on the phase diagram in Fig.~\ref{fig:pd_c2n1}.

\begin{figure}
    \includegraphics[width=0.9\linewidth]{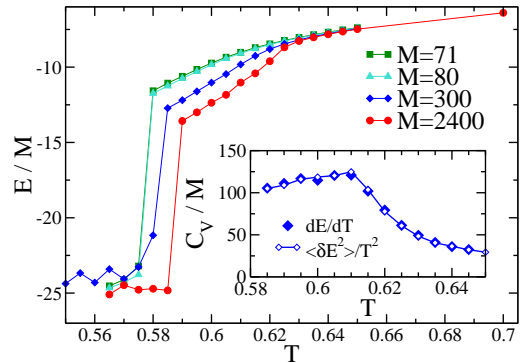}
 	\caption{\label{fig:c2n1_chenr}Average energy per chain for four different C2N1 systems with a packing fraction of approximately 0.097. $E(T)$ data from systems of $M=71$, 80, 300 and 2400 chains are shown, with simulation box sizes of $26^3$, $27^3$, $42^3$ and $84^3$, respectively. Energies are normalized by the number of chains in each system. Inset: Heat capacity per chain, $C_v/M$, for $M=300$ as a function of temperature $T$. As a test of the equilibration of our simulations, $C_v$ is calculated both from a numerical differentiation of $E(T)$ ($dE/dT$) and from the size of the energy fluctuations ($\left <\delta E^2\right>/T^2$) taken separately at each $T$.}
 	
\end{figure}

To further elucidate the phase behavior of the C2N1 sequence, we simulate different-sized systems at a packing fraction of 0.097 (a packing fraction where a clearly evident liquid droplet forms in systems of 300 chains). The size of the simulation box is doubled from $L=42$ to $L=84$, and the number of chains in the system is increased eightfold accordingly. Smaller systems of 71 and 80 chains are also simulated in boxes of length $L=26$ and $L=27$, respectively, which have packing fractions approximately equivalent to the original system (slightly lower and slightly higher). Average energies at each temperature are shown in Fig.~\ref{fig:c2n1_chenr}, normalized by the number of chains in each system.

At high temperature, the average energy per chain is very similar for all system sizes. Here, chains are homogeneously dispersed throughout the simulation boxes; scaling the box size up or down does not affect the average energy per chain. Once liquid droplets begin forming, the energies diverge. As the number of chains in the system is increased, a kink in $E(T)$ corresponding to this transition toward a liquid state becomes more apparent. As the system size is increased, the size of the liquid droplet that forms increases in tandem. Chains within a liquid droplet tend to form lower energy interactions than those at the surface or in the gas state; with a droplet of larger radius, a greater percentage of chains are situated within the interior of the droplet instead of at its the surface, and the lower energy associated to these chains has greater influence on the average energy of the system as a whole. This leads to a sharper kink appearing in the $E(T)$ plot at the temperature where liquid droplets begin forming.
This indicates that the precise location of the heat capacity peak observed in the 300-chain system, where $E(T)$ is smoother, is determined by finite-size effects. 
At low temperature, all system sizes exhibit sudden transitions to a crystal. This crystal transition tends to occur at higher temperature for larger system sizes. This is most likely due to the metastability around the transition. At higher temperature, nucleation of a crystal becomes increasingly rare. A system with more chains would have a proportionally higher probability of crystal nucleation.

In the 2400-chain system, local density calculations can effectively pinpoint the coexistence densities of the liquid droplet and gas phases. With the increased system size, the size of the droplet and gas phases grow as well, and thus surface effects have less impact on the calculation of local densities. The density at the core of the droplet becomes more apparent in the distribution. In the original, 300 chain system, the droplet was too small to identify a corresponding peak in local density. 

\begin{figure}[htb]
    \subfloat[]{\label{fig:big_c2n1a}%
        \includegraphics[width=0.8\linewidth]{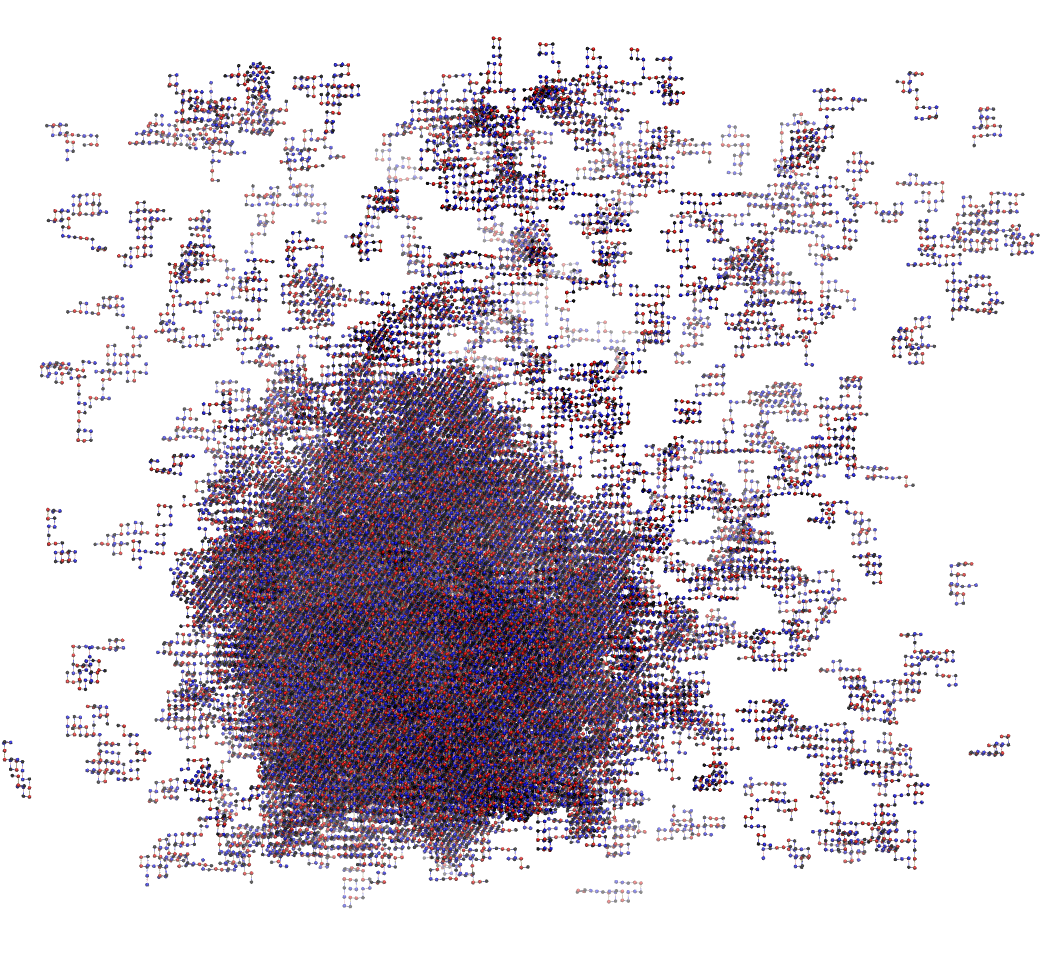}%
    }
    
    \subfloat[]{\label{fig:big_c2n1b}%
        \includegraphics[width=\linewidth]{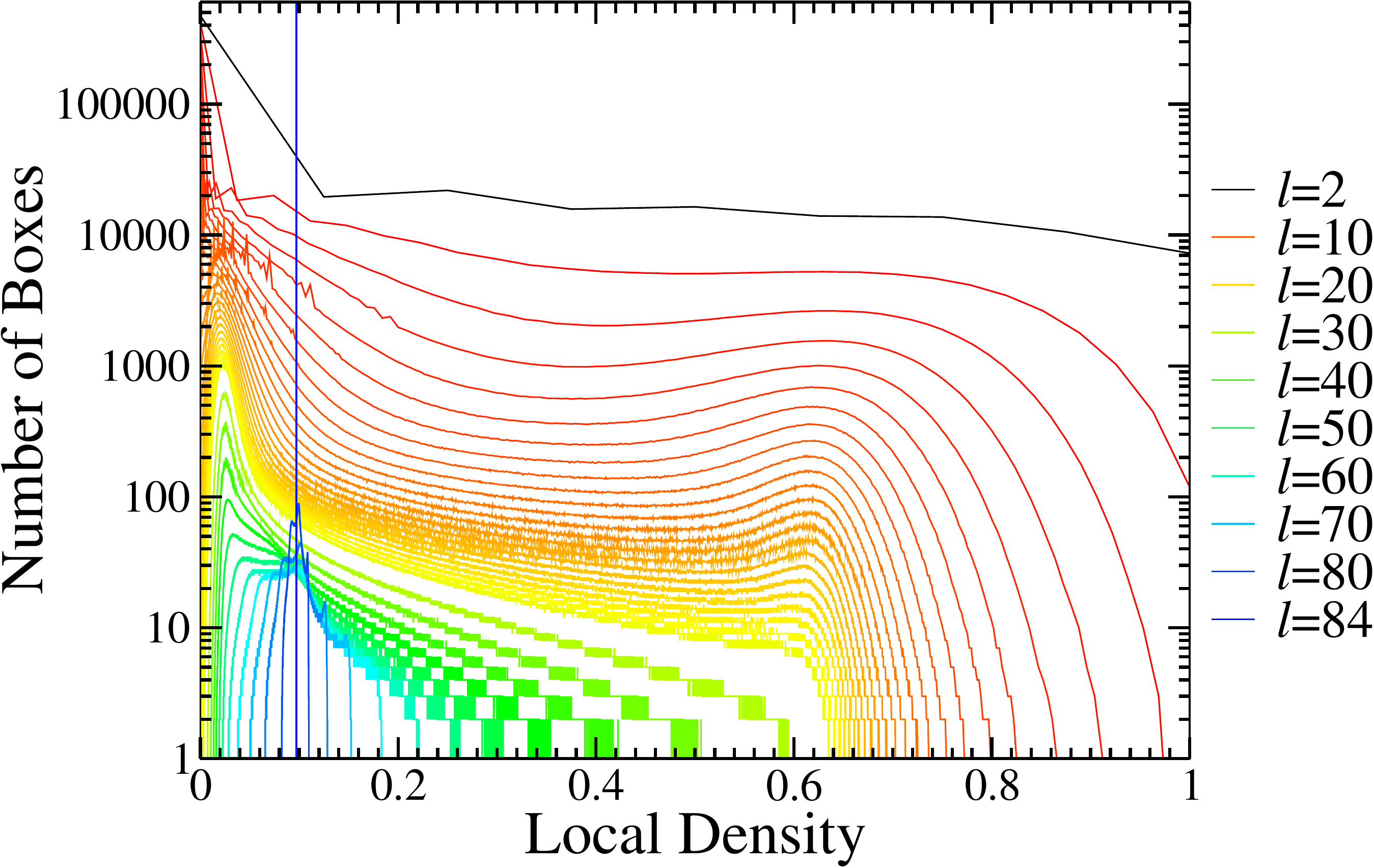}%
    }
 	\caption{\label{fig:big_c2n1}(a) Snapshot of a liquid droplet-containing system of 2400 C2N1 chains at $T=0.6$ and $\Phi=0.097$ ($L=84$). Positive charges are shown in red, negatives charges in blue, and neutrals in black. (b) Local density distribution for the system shown in (a). Colored lines correspond to different box sizes used in the local density calculation, from $l=2$ to $l=84$. Only every fifth box size is shown between $l=25$ and $l=84$. A few selected box sizes are shown in the legend.}
\end{figure}

An image of a liquid droplet and the corresponding local density distribution from a 2400-chain system at $T=0.6$ are shown in Fig.~\ref{fig:big_c2n1}. Local densities are calculated using small box sizes of $l=2$ up to the simulation box size of 84. Over a range of $l$, the distribution is bimodal, with peaks corresponding to the densities of the gas phase and liquid droplet phase. As $l$ is increased, the location of these peaks becomes slightly closer together.  The shift in the density peaks is fairly small: between local box sizes $l=4$ and $l=25$, this shift occurs over a range of approximately 0.05 for the high density peak and 0.005 for the low density peak. The two phases become less distinguishable as local density is probed with increasingly large boxes.  As the box size used in the calculation approaches the system size, the distribution becomes unimodal and approaches the system's packing fraction of 0.097.

\begin{figure}[htbp!]
    \includegraphics[width=\linewidth]{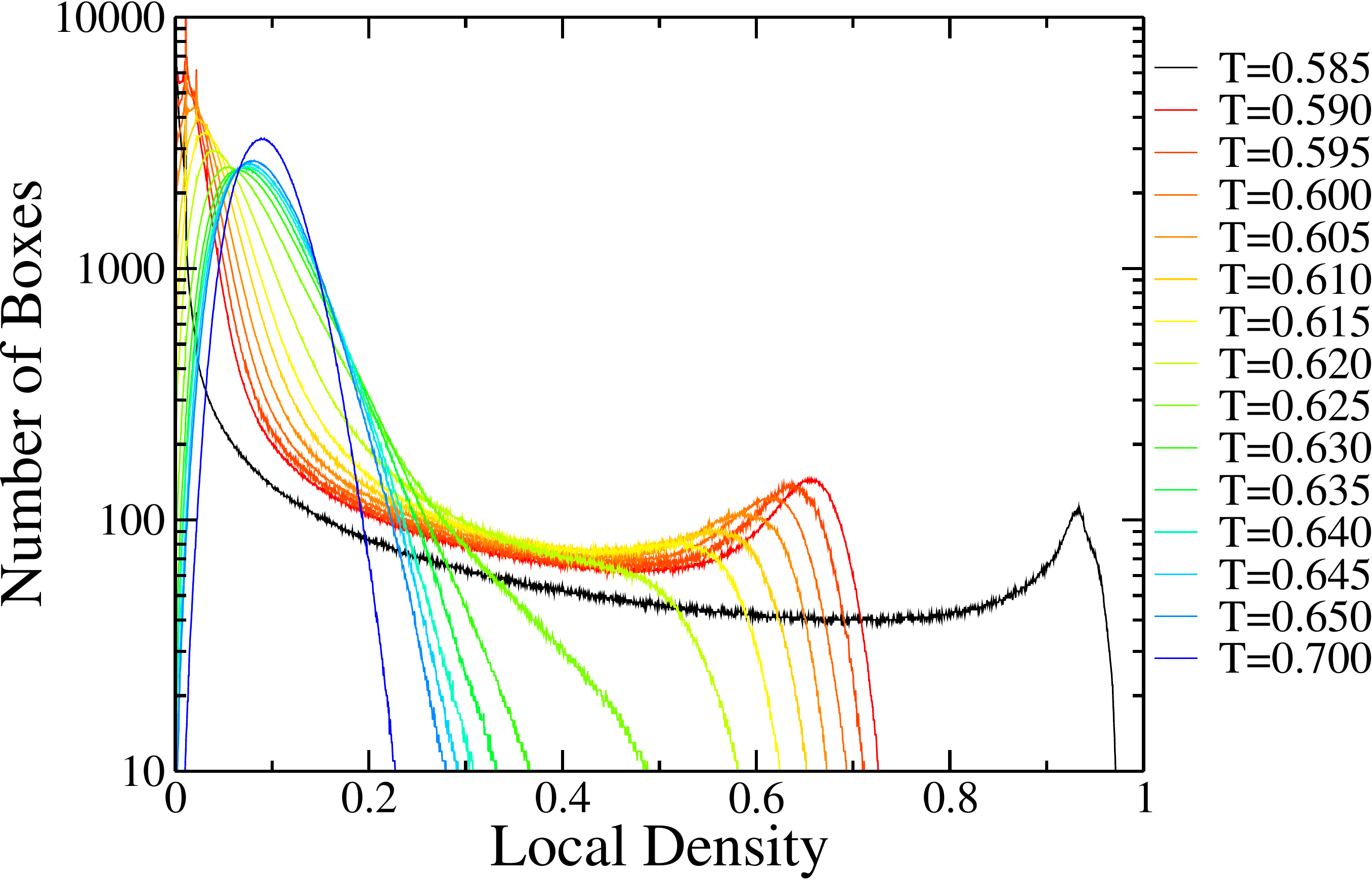}
 	\caption{\label{fig:ld_c2n1}The local density distribution for a chosen local box size $l=13$ at temperatures ranging from $T=0.585$ to $T=0.7$ for systems of 2400 C2N1 chains. As temperature decreases, two peaks appear in the distribution. At $T=0.585$ a crystal has formed.}
\end{figure}

To estimate the coexistence densities of the liquid and gas phases, a local box size of $l=13$ is chosen and the corresponding local density distribution over a range of temperatures for the 2400 chain C2N1 systems is plotted (Fig.~\ref{fig:ld_c2n1}). As temperature is increased, the peaks in the distribution come closer together, until eventually the high density peak vanishes. The distribution becomes unimodal, centred on the system's packing fraction of 0.097, corresponding to a homogeneous fluid system. The highest temperature where the local density distribution is clearly bimodal is 0.615, although the distribution remains broad until 0.63, suggesting the system is still quite heterogeneous until that temperature. This is in agreement with the $E(T)$ data shown in Fig.~\ref{fig:c2n1_chenr}, where the kink in energy corresponding to the onset of liquid droplet formation occurs around $T=0.625$. The local density distribution for a system containing a crystal at $T=0.585$ is also shown in Fig.~\ref{fig:ld_c2n1}. The density of the crystal is clearly distinct from the liquid density, with a value near unity. The peaks in the local density distribution at each temperature above the crystal transition are plotted in the phase diagram shown in Fig.~\ref{fig:pd_c2n1}.

As depicted in the phase diagram (Fig.~\ref{fig:pd_c2n1}), C2N1 chains inhabit a homogeneous fluid state at high temperature. At lower temperatures, systems phase separate into a liquid state and a rare fluid, or gas, state. The location of this liquid-gas coexistence region is approximated by the local density peaks from Figure \ref{fig:ld_c2n1}. From these maxima, a fit was generated using the following procedure. First, the critical temperature $T_C$ is estimated by fitting the density difference of the liquid and gas phases to a scaling law~\cite{frenkel2002,noro2001,vega1992} 
\begin{equation}
    \rho_l - \rho_g = A|T - T_C|^{\beta_c},
\end{equation}
where $\beta_c$ is the critical exponent and $A$ is a constant determined from the fit. Given the finite size of the system and that the coexistence densities are approximate, we also allow $\beta_c$ to vary as a fitting parameter. Then, the critical density, $\rho_C$, is estimated using the law of rectilinear diameters~\cite{frenkel2002,noro2001,vega1992}
\begin{equation}
    \frac{\rho_l-\rho_g}{2} = \rho_C + B|T-T_C|,
\end{equation}
where $B$ is a constant determined from the fit. The critical parameters generated by this fit are $T_C=0.626$, $\rho_C=0.259$, and $\beta_C=0.254$. Fixing $\beta_C=0.325$, the value expected from the 3D Ising model universality class~\cite{Pelissetto2002,panagiotopoulos}, yields $T_C=0.632$ and $\rho_C=0.247$, similar to the values we find from the unconstrained fit of the data. The coexistence curve obtained from these fits is consistent with the energy data for the systems shown in Figure \ref{fig:c2n1_chenr}, which have a packing fraction of 0.097, where the kinks in energy corresponding to a shift from a homogeneous state to a liquid-gas phase separated state occur around 0.625.

Four thresholds in $S(q_{low})$ are displayed on the phase diagram. These mark thresholds in the heterogeneity and compressibility of systems, and roughly approximate a region where a well-formed liquid droplet is present. At higher packing fractions, systems can transition into a liquid-like state without becoming overtly heterogeneous. The liquid state incorporates more of the system and the gas becomes increasingly rare. Therefore, as the system density approaches the density of the liquid state, the increase in $S(q_{low})$ becomes less prominent, and these thresholds no longer detect the occurrence of a homogeneous fluid to liquid-gas transition. As a result, the regions outlined by these thresholds in $S(q_{low})$ do not extend to as high packing fraction as the liquid coexistence densities given by local density calculations. Nevertheless, these thresholds in $S(q_{low})$ suggest that a peak in compressibility occurs near the critical density.

The percolation line, determined from the packing fraction at which the percolation fraction equals 50\% at each temperature,  is also plotted (Fig.~\ref{fig:pd_c2n1}).
This line is fairly vertical, slanting slightly toward higher packing fractions as temperature is decreased. This slanting of the line is the result of the liquid state becoming more condensed at lower \textit{T}.
At packing fractions near 0.1, lowering temperature causes the droplet to shift from percolated to unpercolated. This percolation line emanates from the liquid-gas coexistence region to the left of the critical density, as expected for such a phase transition~\cite{tartaglia2005}.

The fluid-crystal coexistence curve plotted in Figure \ref{fig:pd_c2n1} is very approximate due to the high degree of metastability near the transition (as exemplified in Figure \ref{fig:meta_c2n1}). The points plotted mark the highest temperature state point at each packing fraction where crystallization occurs with systems of 300 chains. Given infinite simulation time or repeated simulations around these temperatures, this line would become smooth and shift to higher temperature. In simulations of 2400 chains, the highest temperature where a crystal forms is 0.595, appearing in only one simulation and after over 500 billion Monte Carlo steps. Since only one packing fraction is simulated with 2400 chains, this state point is not included on the phase diagram for the sake of consistency.

\subsection{C2N3, C2N4 \& C2N16:\\*Sterically Limited Clustering}

The last three sequences studied in this project display another class of phase behavior. As temperature is decreased, these sequences gradually collapse into small clusters or filaments rather than exhibiting an abrupt transition to a condensed phase. At high packing fractions, the low-temperature structures resemble gels, or system-spanning networks. These three sequences, C2N3, C2N4 and C2N16, have larger neutral patches separating pairs of charges and thus lower charge densities than C2N0, C2N1 and C2N2.

At lower temperatures, chains tend to collapse into low-energy structures that optimize the bonds between charges. Neutral beads, meanwhile, have no tendency to form bonds, and can instead remain in a disordered state without impacting the energy of the system. This results in low-temperature, low-energy structures where neutral beads tend to be freely exposed on the surface of clusters while charges interact on the inside, as already seen for the sequences discussed previously. For chains with larger neutral patches in their sequences, this `envelope' of surrounding neutrals sterically limits the interactions of charges and the growth of clusters. As a result, there is less of a drive to form a condensed, solid state. This is what is seen for the C2N3, C2N4 and C2N16. Instead of abruptly transitioning to a solid phase at low temperatures, chains gradually collapse into distinctly shaped clusters through bond-driven interactions. The behavior of these sequences can be identified through energy and heat capacity data.

\begin{figure}
    \subfloat[]{\label{fig:c2n3416a}%
        \includegraphics[width=0.85\linewidth]{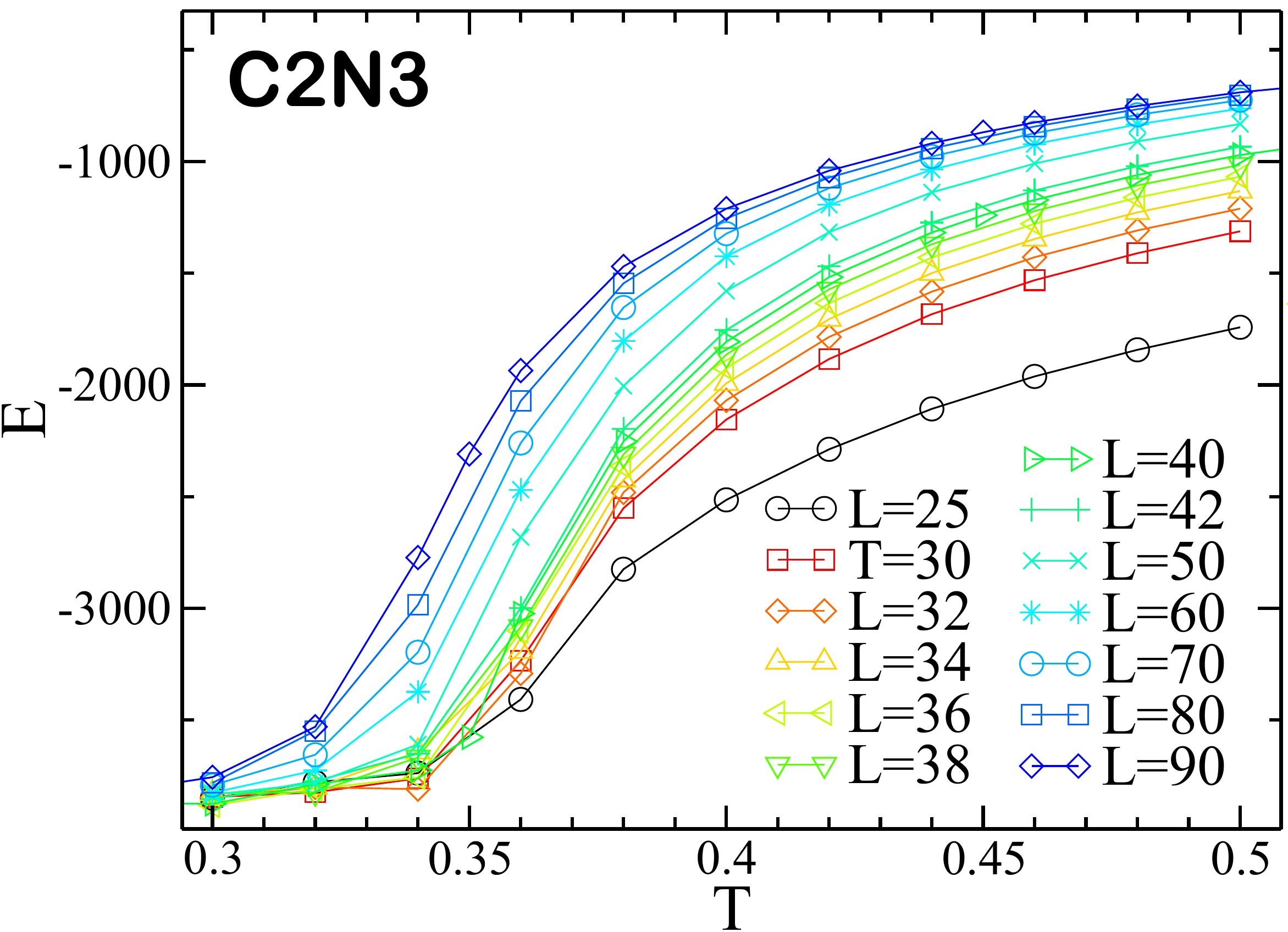}%
    }
    
    \subfloat[]{\label{fig:c2n3416b}%
        \includegraphics[width=0.85\linewidth]{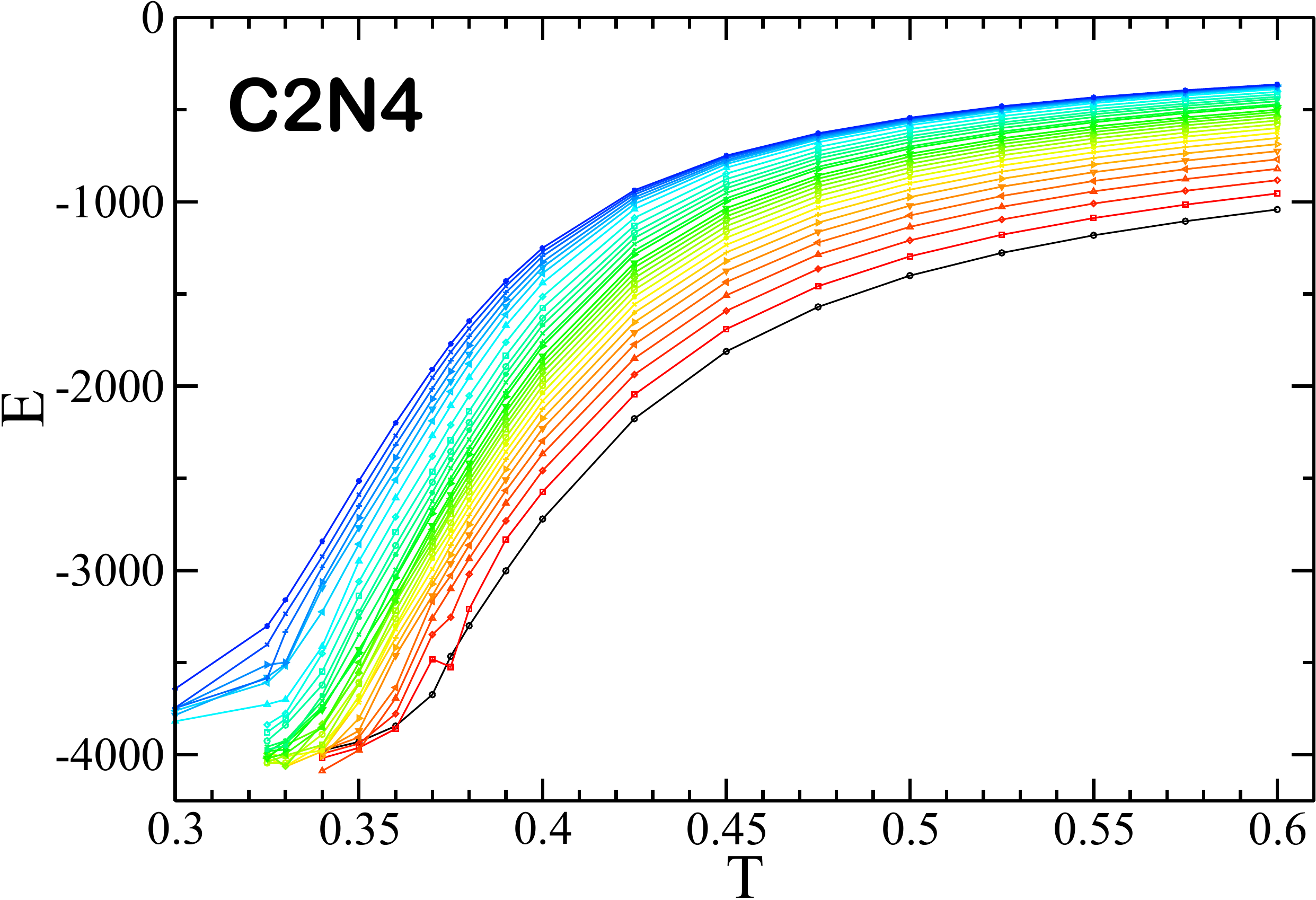}%
    }
    
    \subfloat[]{\label{fig:c2n3416c}%
        \includegraphics[width=0.90\linewidth]{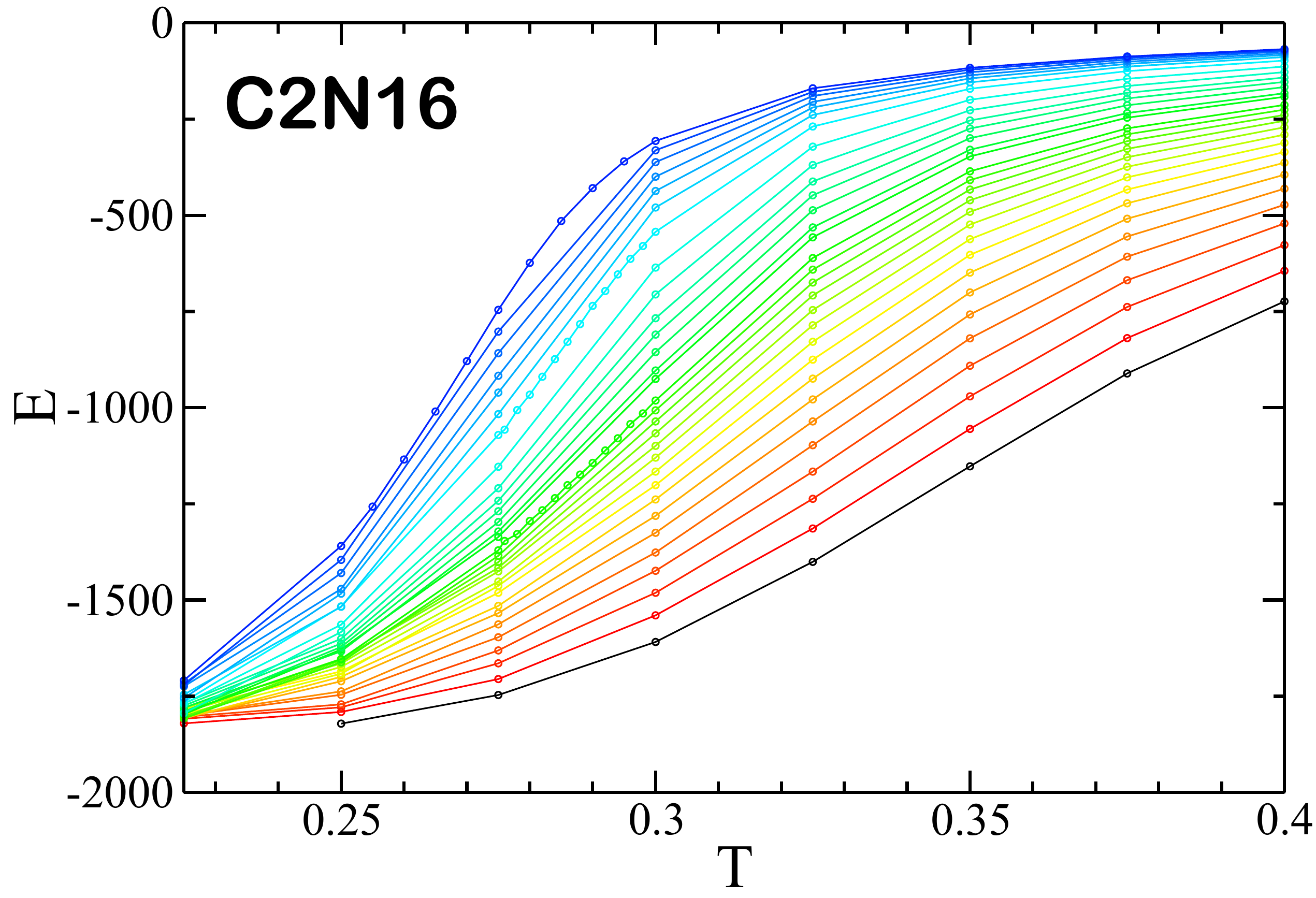}%
    }
 	\caption{\label{fig:c2n3416_enr}Energy versus temperature plots for systems of 300 chains with the (a) C2N3, (b) C2N4 and (c) C2N16 sequence. Lines represent different packing fractions ranging from $\Phi=0.0099$ ($L=90$) to $\Phi=0.461$ ($L=25$). The legend applies to (a); the colors in (b) and (c) have the same meaning as those shown in the legend of Fig.~\ref{fig:c2n20_enr}.}
\end{figure}

These three sequences are simulated in the range of $T=0.100$ to $T=0.600$ with systems of 300 chains. The average system energy of simulations for each temperature and packing fraction is shown in Fig.~\ref{fig:c2n3416_enr}. No abrupt shifts in energy occur for these sequences; instead, there is a gradual lowering of energy as temperature is decreased. The high-energy, homogeneous fluid coagulates into low-energy clusters, forming a gel-like network at higher packing fractions.

\begin{figure}
    \includegraphics[width=\linewidth]{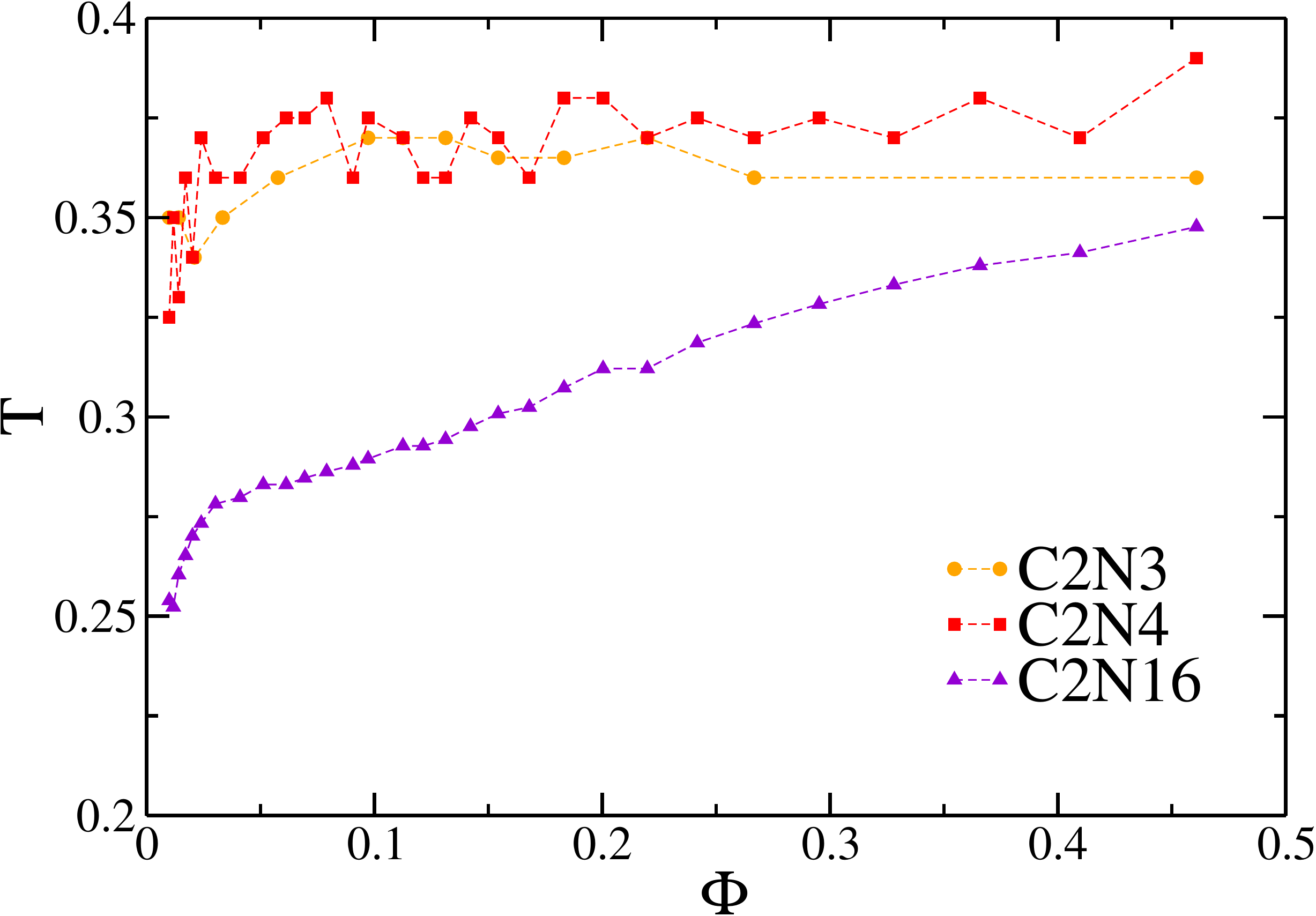}
	\caption{\label{fig:c2n3416_cv}The heat capacity maxima as a function of packing fraction for the C2N3, C2N4 and C2N16 sequences. The behavior of C2N16 has a much higher dependence on packing fraction than the other two sequences.}
\end{figure}

Heat capacity calculations can help characterize the temperature-dependent behavior of these sequences.  For each packing fraction, the heat capacity reaches a maximum at a specific temperature. This maximum corresponds to where the energy of the system is changing the most rapidly with $T$. In other words, this temperature is where the system-wide collapse toward low-energy structures is the most pronounced. Therefore, these maxima can be used to approximate where the shift between the homogeneous fluid and the collapsed state occurs. The heat capacity maxima for all three sequences are plotted in Figure \ref{fig:c2n3416_cv}. These maxima are approximated from the slope of $E(T)$ curves (Fig.~\ref{fig:c2n3416_enr}).

The maxima tend to increase to higher temperatures as packing fraction is increased, most prominently for C2N16. The maxima for C2N3 and C2N4 are fairly flat, occurring around $T=0.350$. The C2N16 maxima are at lower temperature and have a higher dependence on packing fraction. Since C2N16 has a much lower charge density than the other sequences, it is reasonable that it would collapse at lower temperatures.

These three sequences form distinct low-energy structures (Fig.~\ref{fig:dim}). C2N3 tends to form long, filament-like structures. Lines of favorably interacting charges make up the interior of the filament, with an envelope of neutrals surrounding them. These filaments were generally two to three rows of charges wide and could extend through the entire simulation box. By simply stacking on top of each other, chains folded into the optimal energy single-chain conformation for C2N3 shown in Fig.~\ref{fig:minch} could act as building blocks for this structure. In simulations, filaments join and branch off, forming a system-spanning network of linear strands running parallel or perpendicular to each other. The structure is dynamic -- filaments grow, combine and split as simulations progress. There is high variation in their length and width. This network of chains could be described as a gel.

For C2N4 and C2N16, chains tend to collapse into distinct, smallish clusters at lower temperatures. C2N4 forms oblong clusters that usually contain 10 to 20 chains. The core of the cluster is composed of alternating positive and negative charges, with neutral beads surrounding this charged core. C2N16 forms smaller, micelle-like clusters, again with charges isolated at the interior. These clusters do not grow in size as simulations progress; they can dynamically form and dissolve, but their overall size is limited. At higher packing fractions, clusters begin to come into contact, and the size of individual clusters become less well defined. Nevertheless, the overall shape of these micelle-like structures remains, with small cores of charges surrounded by neutral envelopes. However, the chains forming these aggregations become increasingly convoluted. For example, a single C2N16 chain can extend from one charge core to another, with its charge pairs contributing to different cores.

\section{Discussion and conclusions}

The chains we study in this work display varied, sequence-dependent phase behavior.
At high temperature, all sequences tend to disperse uniformly throughout the system. C2N1, with its single neutral beads separating charge pairs, exhibits liquid behavior at an intermediate range of temperature and packing fraction. At low temperatures, C2N0, C2N1, and C2N2 form dense, organized crystals (membrane-like sheets in the case of C2N2). The three sequences with lower charge densities, C2N3, C2N4, and C2N16, do not form a crystal. Instead, as temperature is decreased, they gradually collapse into low-energy states that are condensed at a local level. C2N3 forms a gel-like network of bonded chains. C2N4 and C2N16 form small distinct clusters that display gel-like behavior at high packing fractions. These six sequences are differentiated by the length of the neutral patches between charge pairs. From these findings, it can be concluded that two main factors are responsible for the differences in their phase behavior: the charge density of the chains and the energetic frustration inherent in their sequences, as we discuss below.

For C2N0 and C2N2, individual chains can readily fold into stable, low-energy building blocks for crystal growth (as shown in Fig.~\ref{fig:c2n20_closeup}). These chains are energetically `content'; a single chain can easily fold into an optimal structure, and these folded chains agglomerate to create a crystal. A single C2N1 chain does not individually fold into a structure suitable for crystal formation. The single neutral beads between charge pairs prevent the chain from folding back on itself in a way that brings the opposite charges into full contact. 
Instead, complex inter-chain interactions are required to build a condensed crystal assembly; neighboring chains must fold into highly specific conformations. Because of this, the C2N1 sequence can be described as energetically frustrated -- it has a higher degree of disorder and must exhaustively search through conformational space to find a favorable, low-energy condensed state.  This complexity in C2N1's crystal structure may be the thermodynamic driving force that leads to the stability of an intermediate, liquid-like droplet state above the crystal transition.

As the length of neutral linkers between charges is increased, the tendency for chains to cluster together and to phase separate decreases, with no crystal formation observed at the lowest charge densities. The types of low-energy, low-temperature structures that form for these sequences are highly diverse (Fig.~\ref{fig:dim}). C2N0 and C2N1 form large, globular crystals, which span the entire length of the simulation box. C2N2 forms a flat, membrane-shaped structure, with charges sandwiched between layers of neutral beads. C2N3 forms long filaments, with a charged core insulated by a neutral envelope. C2N16 forms small, micelle-like clusters that never grow in size. The structure of C2N4 clusters is somewhat intermediate between C2N3 and C2N16, forming oblong, neutral-enveloped clusters. As the length of the neutral patches in the sequences increases, these low-energy structures reduce from three-dimensional, to two-dimensional, to one-dimensional filaments, to the zero-dimensional, point-like structures seen for C2N16. This effect is due to the increasingly obstructive effect the neutral patches have on charge interactions. The neutral beads sterically limit the directions from which charges can combine to form a cluster. To optimize charge interactions, neutral beads are pushed to the surface of clusters, leaving charges to interact in the interior. As the length of the neutral patches is increased, the growth of the internal, charged core in these structures is increasingly limited. This results in a low-temperature state with walls of neutrals in C2N2, a ring of neutrals in C2N3, and a sphere of neutrals surrounding the charged core in C2N16. The types of low-energy structures that are seen for these sequences and the driving forces for their formation can relate to real protein behavior.
The effects on liquid-gas phase separation and reduction in dimensionality are similar to what is seen in charged colloids~\cite{tartaglia2004,sciortino2005,zaccarelli2005} and attractive colloids with reduced valence~\cite{tartaglia2005,sciortino2007,liu2007}.

Intrinsically disordered proteins are thought to drive the formation of membrane-less organelles through electrostatically driven phase separation. These intracellular condensates made up of proteins and RNA often exhibit liquid-like behavior. Some membrane-less organelles have more gel-like properties. Some can condense into solid, fibrillar aggregates under certain cellular conditions. The formation of dense, pathological protein aggregates in cells, such as amyloid fibrils, is a hallmark of many diseases~\cite{chiti2017}. Metastable transitions between liquid-like condensates made up of IDPs and more solid-like states have been identified, which could be of biomedical importance in studying these diseased states~\cite{shin}. With all this in mind, the phase behavior of the charged, linear chains studied in this project reveal important connections to intracellular behavior. As would be expected, our findings show that sequences with higher charge density have a higher tendency to crystallize. Ordered, even, low-complexity sequences like C2N0 and C2N2 crystallize over a larger range of conditions. For all sequences, there is an increased tendency to crystallize, gelate or coagulate with lower temperature and higher concentration (packing fraction), as seen for real protein systems. C2N3, C2N4 and C2N16 systems collapse into dense filaments or gel-like states in a temperature and concentration-dependent manner, emulating pathological protein behavior. C2N1, with its energetically frustrated structure, exhibits liquid behavior that maintains metastability with respect to a solid at lower temperatures. The behavior of the C2N1 sequence directly connects to IDP behavior. IDPs are proteins that cannot stably fold into a low energy structure; in isolation, they are dynamic and flexible. Similarly, C2N1 has a reduced tendency to fold uniquely relative to the other sequences with similar charge density. Its sequence gives rise to increased conformational disorder; in other words, C2N1 is intrinsically disordered. Under certain cellular conditions, IDPs form specific multivalent interactions that can lead to the formation of a liquid-like assembly. This is what is seen for the C2N1 sequence: it agglomerates into a liquid-like state over a specific range of conditions. In more extreme conditions, C2N1 chains condense into solid agglomerates with a high degree of metastability at the transition, as has been seen for real biomolecular condensates. Our findings suggest that multiple sequence properties influence the phase behavior and the formation of liquid-like states for charged polymers like IDPs. In the case of C2N1, we see properties that are conducive to liquid formation. On the one hand, the C2N1 sequence has sufficient charge density to form a condensed phase through multiple inter-chain bonds. However, unlike other sequences that can quickly condense from a fluid state to a crystal, C2N1's charge pattern leads to the presence of an intermediate liquid state. The configurations of C2N1 chains that give rise to a crystal are significantly different from the optimally folded state of an isolated C2N1 chain. Its sequence results in the need for elaborate inter-chain bonding to build the crystal structure, and thus the stability of a liquid phase.

Our analysis could be extended to other sequences to determine more general properties of the phase behavior for these types of systems. With C2N1 in mind, new sequences could be envisioned that would likely exhibit liquid-like behavior. To begin with, the C2N1 sequence could be carefully tweaked (for instance, by adding or removing neutral beads) without drastically affecting its folding behavior. Such an approach could modulate the phase behavior of the sequence, changing the shape and location of the liquid-gas coexistence curve. Looking beyond C2N1, sequences with similar charge density and with a similarly low propensity to fold into simple, regular, condensed low-energy structures could likely exhibit liquid-like behaviour before crystallization. 

The model used here could be adjusted in future work. The cut-off length of interactions between charges could be increased, which may help identify the role of long-range interactions between charged regions of chains.  Moreover, long-range charge-charge interactions would be needed in order to study the effects of a net charge on the chains. In the present study, neutral beads and empty lattice sites do not interact energetically with other beads; other types of interactions could be included.  Several different types of interactions are thought to contribute to IDP phase separation, including  $\pi$-$\pi$ stacking (i.e., interactions between aromatic residues), cation-$\pi$, dipole-dipole interactions and hydrogen bonding~\cite{brangwynne,Murthy2019}. Hydrophobic attraction is a key stabilizing force in protein folding~\cite{Kauzmann1959} and aggregation~\cite{Mahmoudinobar2019}, and it can also play a role in liquid-liquid phase separation~\cite{Rauscher2017}. The strong temperature dependence of the hydrophobic effect, which underlie, e.g., the unfolding of some proteins at low temperature~\cite{Dias2008}, might therefore influence the phase behavior of IDPs. Indeed, an explicit temperature dependence on the strength of interactions were necessary to include in coarse-grained models of liquid-liquid phase separation in order to explain the existence of a lower critical solution temperature (LCST) exhibited by some IDPs, i.e., phase separation occurring upon an increase in temperature~\cite{Dignon2019}. Including additional effects in our model, and consequently more amino acid types, would allow for the simulation of real IDP sequences with experimentally measured phase behaviors.

%, but its role in LLPS is still unclear. While the LLPS of the Alzheimer's disease related protein tau is not strongly dependent on hydrophobic residues~\cite{Boyko}, the LLPS of elastic-like peptides are driven by hydrophobic attractions~\cite{Pomes}.
%Removed: Sequences with more types of beads would allow for the modeling of hydrophobic, dipole and other interactions.  

Much remains to be elucidated about the physics behind charged polymers and their phase separation. The relationship between sequence properties and phase behavior is complex and multi-faceted. Charge density, the length of neutral patches, and the intrinsic disorder of chains play central roles. To develop an understanding of this interplay, systematically-varied sets of representative sequences must be studied. This work serves to characterize the diverse behavior of six such sequences.

\begin{acknowledgments}
We acknowledge the work of Daniel Trotter and Dr.~Anders Irb\"ack in the development of early versions of the computer code. This research was enabled in part by support provided by Compute Canada (www.computecanada.ca). S.~W. and I.~S.-V. acknowledge financial support from NSERC (Canada).
\end{acknowledgments}

%\bibliography{latmod}% Produces the bibliography via BibTeX.

%

\end{document}